\documentclass[twocolumn]{aastex631}

\newcommand{\HI}{H\textsc{i}}
\newcommand{\Msolar}{M$_{\odot}$}
\newcommand{\kms}{km\,s$^{-1}$}

\shorttitle{Dark clouds in the Leo I Group}
\shortauthors{R. Taylor et al.}

\begin{document}

\title{The Arecibo Galaxy Environment Survey XII : Optically dark HI clouds in the Leo I Group}

\correspondingauthor{Rhys Taylor}
\email{rhysyt@gmail.com}

\author[0000-0002-3782-1457]{Rhys Taylor}
\affiliation{Astronomical Institute of the Czech Academy of Sciences,\\
Bocni II 1401/1a, \\
141 00 Praha 4,
Czech Republic}

\author{Joachim K\"{o}ppen}
\affiliation{Astronomical Institute of the Czech Academy of Sciences,\\
Bocni II 1401/1a, \\
141 00 Praha 4,
Czech Republic}
\affiliation{Institut f\"{u}r Theoretische Physik und Astrophysik der Universit\"{a}t zu Kiel,\\
D-24098, Kiel, Germany.}

\author[0000-0002-1640-5657]{Pavel J\'{a}chym}
\affiliation{Astronomical Institute of the Czech Academy of Sciences,\\
Bocni II 1401/1a, \\
141 00 Praha 4,
Czech Republic}

\author[0000-0002-1261-6641]{Robert Minchin}
\affiliation{Stratospheric Observatory for Infrared Astronomy/USRA,\\
NASA Ames Research Center, MS 232-12,\\
Moffett Field, CA 94035, USA}

\author[0000-0001-6729-2851]{Jan Palou\v{s}}
\affiliation{Astronomical Institute of the Czech Academy of Sciences,\\
Bocni II 1401/1a, \\
141 00 Praha 4,
Czech Republic}

\author[0000-0002-5993-9069]{Jessica L. Rosenberg}
\affiliation{George Mason University,\\
4400 University Drive, MS 3F3,\\
Fairfax, VA 22030,\\
USA}

\author{Stephen Schneider}
\affiliation{UMass Department of Astronomy,\\
517K Lederle Graduate Research Tower,\\
Amherst, MA 01003,\\
USA}

\author[0000-0003-1848-8967]{Richard W\"{u}nsch}
\affiliation{Astronomical Institute of the Czech Academy of Sciences,\\
Bocni II 1401/1a, \\
141 00 Praha 4,
Czech Republic}

\author[0000-0002-7898-5490]{Boris Deshev}
\affiliation{Astronomical Institute of the Czech Academy of Sciences,\\
	Bocni II 1401/1a, \\
	141 00 Praha 4,
	Czech Republic}

\begin{abstract}
Using data from the Arecibo Galaxy Environment Survey, we report the discovery of five \HI{} clouds in the Leo I group without detected optical counterparts. Three of the clouds are found midway between M96 and M95, one is only 10$^{\prime}$ from the south-east side of the well-known Leo Ring, and the fifth is relatively isolated. \HI{} masses range from 2.6$\times$10$^{6}$\,--\,9.0$\times$10$^{6}$\Msolar{} and velocity widths (W50) from 16\,--\,42\,\kms{}. Although a tidal origin is the most obvious explanation, this formation mechanism faces several challenges. For the most isolated cloud, the difficulties are its distance from neighbouring galaxies and the lack of any signs of disturbance in the \HI{} discs of those systems. Some of the clouds also appear to follow the baryonic Tully-Fisher relation between mass and velocity width for normal, stable galaxies which is not expected if they are tidal in origin. Three clouds are found between M96 and M95 which have no optical counterparts, but have otherwise similar properties and location to the optically detected galaxy LeG 13. While overall we favour a tidal debris scenario to explain the clouds, we cannot rule out a primordial origin. If the clouds were produced in the same event that gave rise to the Leo Ring, they may provide important constraints on any model attempting to explain that structure.
\end{abstract}

\section{Introduction}
The Leo I group is well-known for its numerous optically dark \HI{} features. At approximately 11 Mpc distance, the group consists of at least 4 major galaxies. Its most unusual feature is the spectacular Leo Ring, a 225 kpc diameter arc of approximately 1.7$\times$10$^{9}$\Msolar{} of \HI{} (\citealt{schnleo}, \citealt{aaleo}). Some isolated patches of optical emission have been detected at various points within the Ring (\citealt{thileo}), but on the large scale the \HI{} has no associated optical component. The origin of this structure is still unclear. It has been suggested to be primordial (\citealt{schnleo}, \citealt{leoprim}), though metallicity measurements suggest otherwise (\citealt{rose}, \citealt{leomet}). One alternative explanation is a galaxy-galaxy collision (\citealt{leocollide}), which well reproduces the overall morphology but has difficulties explaining the observed kinematics.

At approximately the same location on the sky as some parts of the Ring, but at a higher velocity, extended \HI{} emission is also seen around the galaxy NGC 3389 (\citealt{aaleo}). This feature, although less dramatic, itself still extends across $\sim$\,90 kpc. It contains several optically bright galaxies, and appears relatively easy to explain by tidal interactions. Elsewhere in Leo, the Leo Triplet contains a $\sim$100 kpc \HI{} plume (\citealt{aaleo}) while even larger features are reported in \cite{aaleoleis}.

Giant, elongated \HI{} structures are often easily explained by galaxy-galaxy interactions (e.g. \citealt{tooms}, \citealt{bekki}, \citealt{duc}, \citealt{me17},) or ram pressure stripping (\citealt{oo05}, \citealt{ton10}, \citealt{me20}). Less common are small, discrete \HI{} clouds with no obvious extended feature to indicate their parent galaxy - we describe a few of these in section \ref{sec:others}, with a longer review given in \cite{me16}. The origin of such clouds tend to be controversial. While they may be the remnants of dispersed streams from tidal encounters, an alternative is that they are dark matter dominated but extremely dim (or even entirely dark) galaxies, either primordial (\citealt{d04}, \citealt{m07}) or resulting from interactions (\citealt{fadug}).

We tested these explanations in \cite{me16} and \cite{me17}. We found that the \textit{majority} of cases of such isolated clouds can indeed be explained as tidal debris, as may particular features embedded within larger structures, such as VIRGOHI21 (\citealt{m07}, \citealt{aavhi21}). In contrast, we also showed that while our simulations readily produced isolated clouds of low velocity widths ($<$\,50 \kms{}), clouds of line width $>$ 100 \kms{} were virtually non-existent in those same simulations. This makes the clouds described in \cite{me12} and \cite{me13} particularly challenging to explain.

In this paper we present the discovery of five new clouds in the Leo I region using a deep Arecibo survey. While a handful of such clouds are known in other groups, this is the first time such clouds have been found in this environment. We discuss the possible origin of the clouds, considering the strengths and weaknesses of different models proposed to explain the formation of the Leo Ring. 

The rest of this paper is organised as follows. In section \ref{sec:obs} we describe the observations, data reduction and source extraction of the \HI{} data used in this analysis. The results are presented in section \ref{sec:results} and analysed in section \ref{sec:analysis}. Finally, we summarise our findings in section \ref{sec:findings}. Throughout, we assume a group distance of 11.1 Mpc following \cite{aaleo}. At this distance, the Arecibo 3.5$^{\prime}$ diameter beam has a physical size of 11.3 kpc. \cite{aaleo} give the group velocity dispersion as 175 \kms{}. Since all of our clouds have systemic velocities within 75 \kms{} of M96, we assume that they are all at the group distance of 11.1 Mpc unless stated otherwise.

\begin{deluxetable*}{c c c c c c c c c c}
\tablecaption{Observed \HI{} parameters for our selected clouds. All coordinates are J2000. Clouds are ordered in ascending declination.}
\label{tab:clouds}
\tablehead{
\colhead{Name} & \colhead{R.A.} & \colhead{Declination} & \colhead{Velocity} & \colhead{W50} & \colhead{W20} & \colhead{F$_{tot}$} & \colhead{M\HI{}} & \colhead{S/N$_{peak}$} & \colhead{S/N$_{tot}$}\\
\colhead{} & \colhead{} & \colhead{} & \colhead{\kms{}} & \colhead{\kms{}} & \colhead{\kms{}} & \colhead{Jy\,\kms{}} & \colhead{\Msolar{}} & \colhead{} & \colhead{}
}
\startdata
  Cloud 1 & 10:44:47.79 & 11:27:34.05 & 960 & 38 & 100 & 0.188 & 5.5E6 & 12.3 & 17.0\\
  Cloud 2 & 10:45:59.27 & 11:30:34.66 & 891 & 33 & 58 & 0.201 & 5.9E6 & 13.5 & 15.7\\
  Cloud 3 & 10:45:36.17 & 11:44:20.90 & 894 & 42 & 73 & 0.275 & 8.0E6 & 13.6 & 19.0\\
  Cloud 4 & 10:44:56.44 & 11:54:52.40 & 879 & 31 & 46 & 0.308 & 9.0E6 & 19.6 & 24.7\\
  Cloud 5 & 10:50:25.91 & 12:09:54.62 & 916 & 16 & 27 & 0.088 & 2.6E6 & 9.2 & 8.2\\
  Cloud 6 & 10:45:00.04 & 13:26:04.37 & 860 & 34 & 45 & 0.172 & 5.0E6 & 9.9 & 11.0\\
\enddata
\end{deluxetable*}

\section{Observations}
\label{sec:obs}
\subsection{The AGES data}
\label{sec:obspartone}
The Arecibo Galaxy Environment Survey AGES, \citealt{auld}) was a blind \HI{} survey performed at the Arecibo radio telescope from 2006-2019. This was part of a tiered approach to \HI{} surveys taking advantage of the then-new ALFA (Arecibo L-band Feed Array) receiver. To this end, ALFALFA (Arecibo Legacy Fast ALFA survey; \citealt{alfalfa}) covered approximately 7,000 square degrees to an \textit{rms} sensitivity of about 2.2 mJy; AGES provides a medium-deep survey of 200 square degrees to 0.7 mJy; and AUDS (Arecibo Ultra Deep Survey, \citealt{auds}) examined 1.4 square degree to 0.08 mJy \textit{rms}). AGES targeted the full range of galaxy environments from voids to clusters, with one of the major goals being the detection of optically dark \HI{} sources that could not be discovered by optical surveys. The 1$\sigma$ column density sensitivity of AGES is N$_{\HI{}}$ = 1.5$\times$10$^{17}$cm$^{-2}$ (0.001 \Msolar{}pc$^{-2}$) for a source which fills the beam at 10 \kms{} velocity resolution (\citealt{olivia}); while AGES has a maximum velocity resolution of 5 \kms{}, we Hanning smooth the data to 10 \kms{} for better sensitivity and reduction of artifacts such as Gibbs ringing.

The AGES observing strategy and data reduction techniques have been described in detail in \cite{auld}, \cite{m10} and \cite{d11}. They are here only summarised. AGES was a fixed-azimuth drift scan survey. The telescope was pointed to the start position of a scan and the sky allowed to drift overhead, typically for 20 minutes (as here) to cover 5 degrees of Right Ascension with the seven beams of the ALFA receiver. Each point took about 12 seconds to cross the beam. When complete, the telescope was re-oriented to the next scan, staggered by one-third of the beam size (3.5$^\prime$) to create a fully Nyquist-sampled map. The total on-source integration time per point was 300 seconds.

Observations of the Leo field were carried out in 2012 (January--February; November--December), 2013 (February, April--May, November--December), 2014 (January), 2015 (January--March, May, December), 2016 (December), 2017 (January--February), 2018 (January--May, December), and 2019 (January). The observations cover a field of approximately 5 degrees of Right Ascension and 4 degrees of declination, centred on 10:45:00, 12:48:00. The full range of the field is 10:34:49\,--\,10:55:14 in Right Ascension and 10:42:58\,--\,14:52:01 in Declination. The bandwidth covered is equivalent in heliocentric velocity from -2,000 to +20,000\,\kms{}. Owing to the hexagonal beam pattern of ALFA, sensitivity is reduced over a few arcminutes near the spatial edges of the field. Sensitivity is also reduced over the final 1,000\,\kms{} of spectral baseline. We are here only presenting observations of the M96 subgroup, which is in a frequency range unaffected by interference.

The data from the two linear polarisations for each of ALFA's seven beams is reduced using the \textsc{livedata} and \textsc{gridzilla} software packages. Following \cite{me14}, we further apply a second-order polynomial to the spectra at every pixel in the resulting data cube, further reducing the impact of continuum sources. Additionally, we apply a \textsc{medmed} (\citealt{put}) correction to each spatial baseline. This divides each baseline into five equal segments, computes the median value of the flux in each one, and then subtracts the median of the medians. The advantage of this is a significant reduction in `shadows' in the data, caused by over-subtraction of bright extended sources (an example is shown in \citealt{m10}).

\subsection{Source extraction}
We here concentrate on the primary target region, the M96 subgroup in Leo. The primary target for AGES in this region was the Leo Ring. In the course of investigating this feature, we serendipitously discovered several discrete \HI{} clouds which are optically undetected (we refer to them as `dark' by convention, see section \ref{sec:otherlambda} for details), at least one of which appears to have no association with the Ring. We restrict our present analysis to these clouds, and leave discussion of the Ring itself, as well as a catalogue over the full AGES spectral bandwidth, to future works. Here we limit our search to the M96 subgroup, within the velocity range 300\,--\,1,600 \kms{}. All of the clouds are within 75 \kms{} of the velocity of M96.

We searched the full spatial field by eye using the FRELLED software described in \cite{me15}. While this has been considerably updated since its initial release, to be described in a future paper, the basic source extraction procedure remains the same (see, e.g. \citealt{me14}). We first display the cube in a volume render, defining masks around anything that visually resembles an \HI{} source. The advantage of the volume rendering technique is that it enables very rapid masking, since the user can see the three-dimensional (velocity space) structure of the source at a glance. The downside is that this often makes the faintest sources hard to detect. We therefore proceed to a second stage. With the masks still hiding the sources already found, we search each individual slice (i.e. channels and position-velocity slices) of the data cube and apply additional masks if any more sources are detected - in this case, no further sources were found. This approach is designed to combine the key advantages of the rapid cataloguing speed enabled through volumetric rendering, together with the sensitivity of more traditional visualisation techniques.

We quantify the \HI{} parameters of each detection using the \textsc{miriad} package \textit{mbspect} (\citealt{miriad}). We define the spectral profile window by eye and \textit{mbspect} applies position fitting by computing the centroid of a 2D-Gaussian to the integrated map over the specified velocity range. It then computes the total integrated flux (S$_{\HI{}}$) and velocity width at 50\% (W50) and 20\% (W20) of the peak flux. The \HI{} mass is computed by the standard equation :
\[M\HI{} = 2.36\times10^{5}\times d^{2}\times S_{\HI{}}\]
We searched for optical counterparts by checking the Sloan Digital Sky Survey (SDSS) and NASA Extragalactic Database (NED) at the position-fitted coordinates of the source, searching within a 1.7$^\prime$ (half the FWHM) radius and checking whether any visible or listed sources have optical redshift measurements within 200 \kms{} of the \HI{} measurement. These search parameters are deliberately very generous to maximise the chance of finding any plausible optical counterpart. Typically, from other AGES data sets (e.g. \citealt{me12}) optical counterparts have spatial positions within about 20$''$ of the \HI{} coordinates; the resolution of the survey is usually more than sufficient for identifying any counterparts where they exist. Similarly the difference between optical and \HI{} redshifts for candidate optical counterparts is usually less than 50 \kms{}.

Quantifying the \HI{} mass sensitivity is complicated as the signal-to-noise (S/N) level for a source of a given mass varies according to both its line width and peak flux : no matter how massive, at a sufficiently high line width a source will always become so faint as to be undetectable. As a rough estimate, a source of 4$\sigma$ with top-hat profile of width 50 \kms{} is a reasonable approximation for the faintest readily detectable source, which at 11.1 Mpc corresponds to an \HI{} mass of 4.1$\times$10$^{6}$\,\Msolar. Some of the sources presented here have masses close to or even below this value, raising the issue as to whether our source extraction is complete for clouds of the given parameters. But a top-hat profile is not realistic. A more sophisticated sensitivity estimate can be provided using the \cite{aasn} definition of integrated S/N, which accounts for the variation in line width and peak flux together. ALFALFA is reckoned to be complete for total S/N values, by this definition, above 6.5. By injecting several thousand artificial sources of different line widths and fluxes into galaxy-free data cubes, and running our own source extraction procedures, we show that AGES completeness well-matches the ALFALFA criteria (Taylor et al. in preparation). All of the clouds presented herein have total S/N values exceeding 6.5, showing that we are indeed complete for clouds similar to those we detect.

\section{Results}
\label{sec:results}
\subsection{AGES \HI{} measurements}
Excluding the Leo Ring, we found a total of 29 sources. Most of these are previously identified sources with clear optical counterparts, generally bright galaxies with matching optical redshift determinations. Here we present five sources which lack any apparent optical counterpart, and, for comparison, one otherwise similar \HI{} detection (cloud 4) which has a likely association with an optically identified galaxy (LeG 13). We include this source in our discussion since it is found in close proximity to three optically dark sources and its \HI{} otherwise strongly resembles those clouds, raising the question as to why this cloud alone should be optically visible (we will refer to it as a `cloud' throughout this work only for convenience, but note that the optical source does not have an optical redshift measurement - see section \ref{sec:otherlambda}).

We present the observational parameters of our six \HI{} clouds in table \ref{tab:clouds}. All the clouds are broadly similar in their \HI{} content, with masses ranging from 2.5\,--\,9.0$\times$10$^{6}$\,\Msolar and velocity widths (W50) from 15\,--\,42\,\kms{}. There is some ambiguity in the velocity widths though, with some of the W20 measurement being up to 100 \kms{}, which we will discuss further in section \ref{sec:btfr}. Table \ref{tab:clouds} also gives the integrated S/N as defined by \cite{aasn}, where a value exceeding 6.5 generally indicates a reliable detection. All clouds exceed this value. Although cloud 5 is only marginally above the threshold, it is easily visible in the data cube.

\begin{figure*}
\centering
\includegraphics[width=180mm]{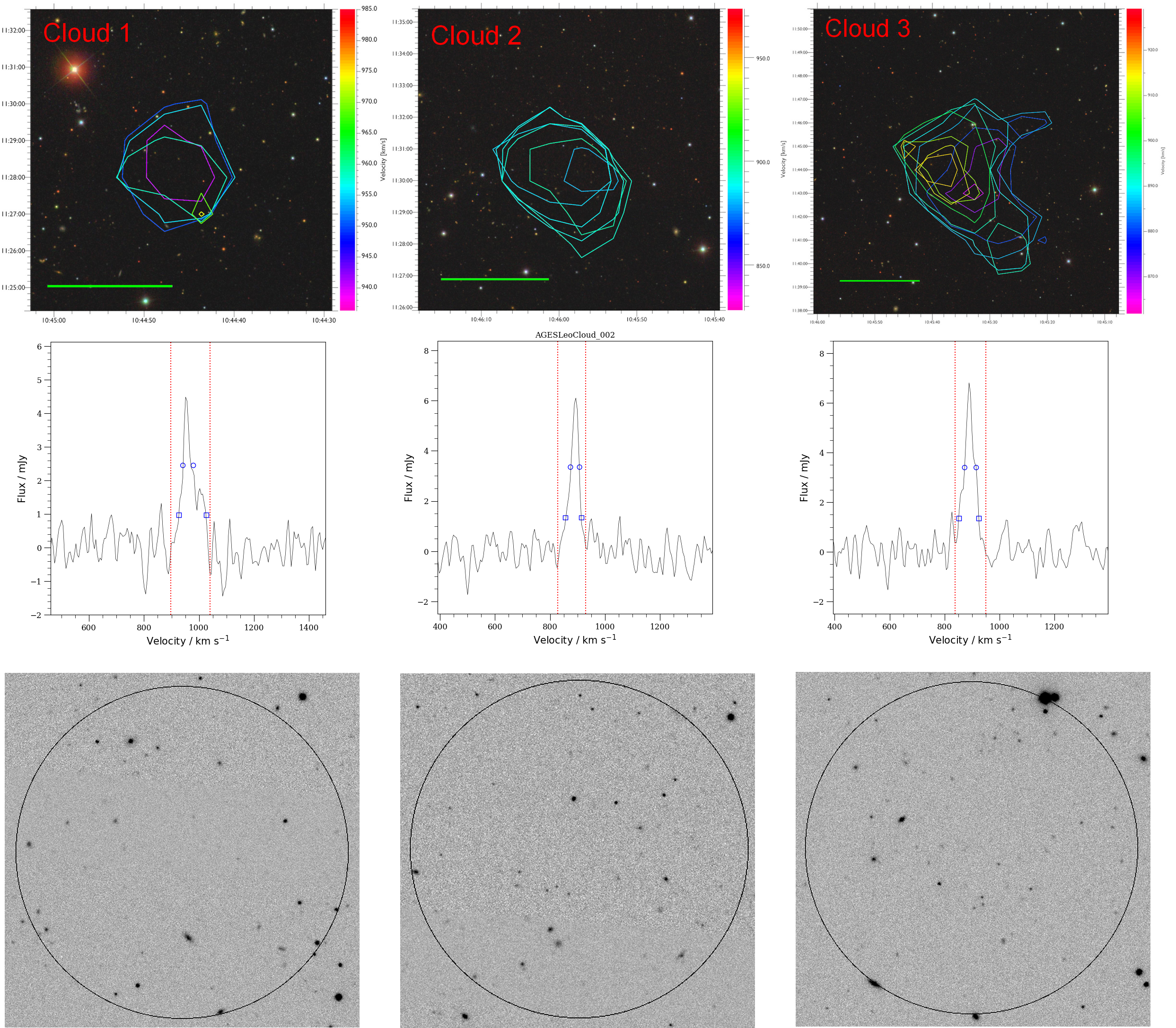}
\caption{Clouds 1-3. The upper panel shows renzograms with the contour at 3.5$\sigma$ and the Arecibo 3.$5^\prime$ beam (equivalent diameter 11.3 kpc) as a green line, overlaid on an a RGB image from the SDSS. The middle panel shows the spectra - red dashed lines show the profile window used for computing the \HI{} parameters, blue circles show the positions of the W50 measurement and blue squares the W20 values. The lower panel shows the stacked $g,r,i$ bands from the SDSS, with the black circle showing the Arecibo beam size, centred on the coordinates of the \HI{} detections.}
\label{fig:cloudset1}
\end{figure*}

\begin{figure*}
\centering
\includegraphics[width=180mm]{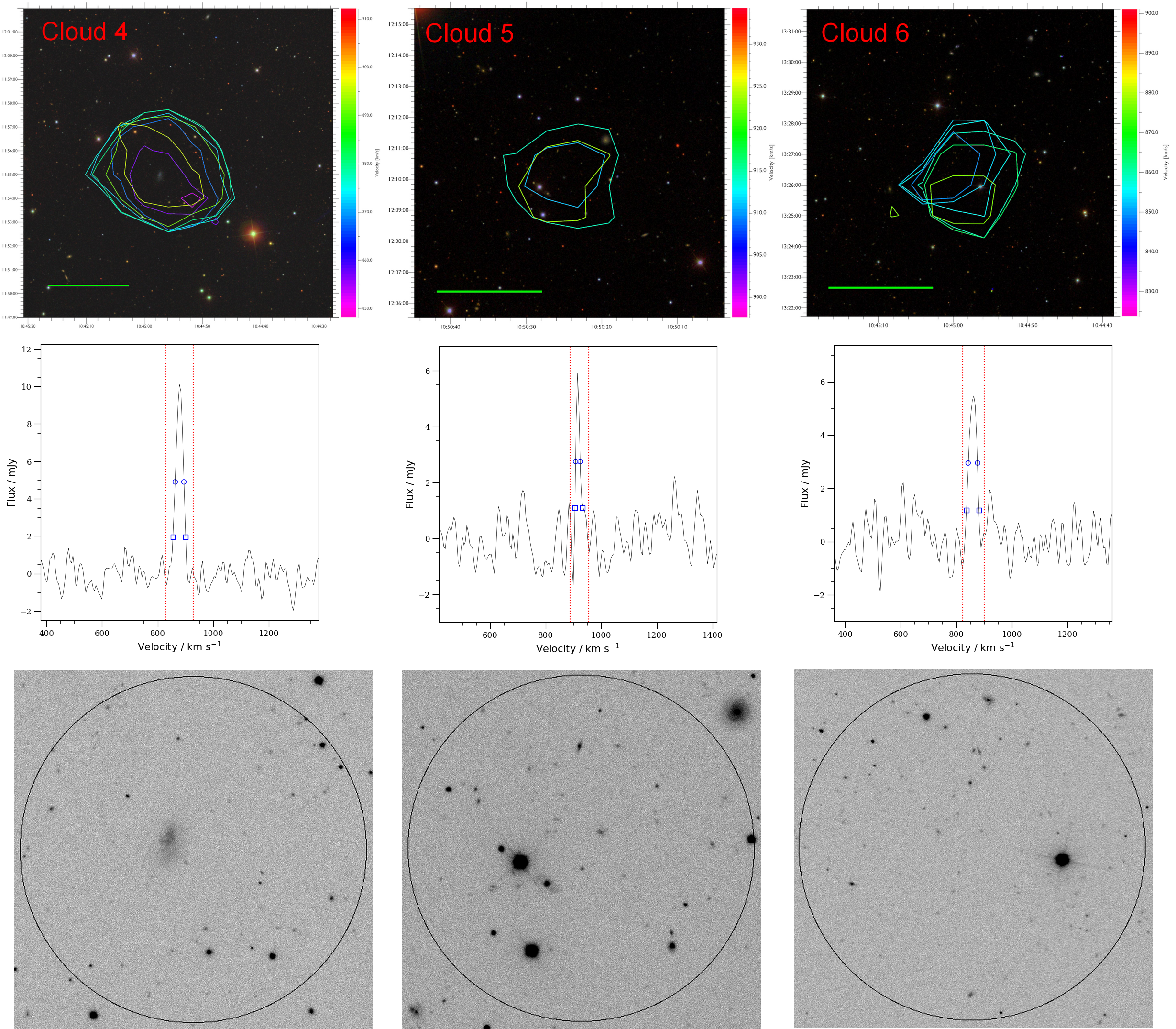}
\caption{As for figure \ref{fig:cloudset1} but for clouds 4-6.}
\label{fig:cloudset2}
\end{figure*}

\begin{figure*}
\centering
\includegraphics[width=180mm]{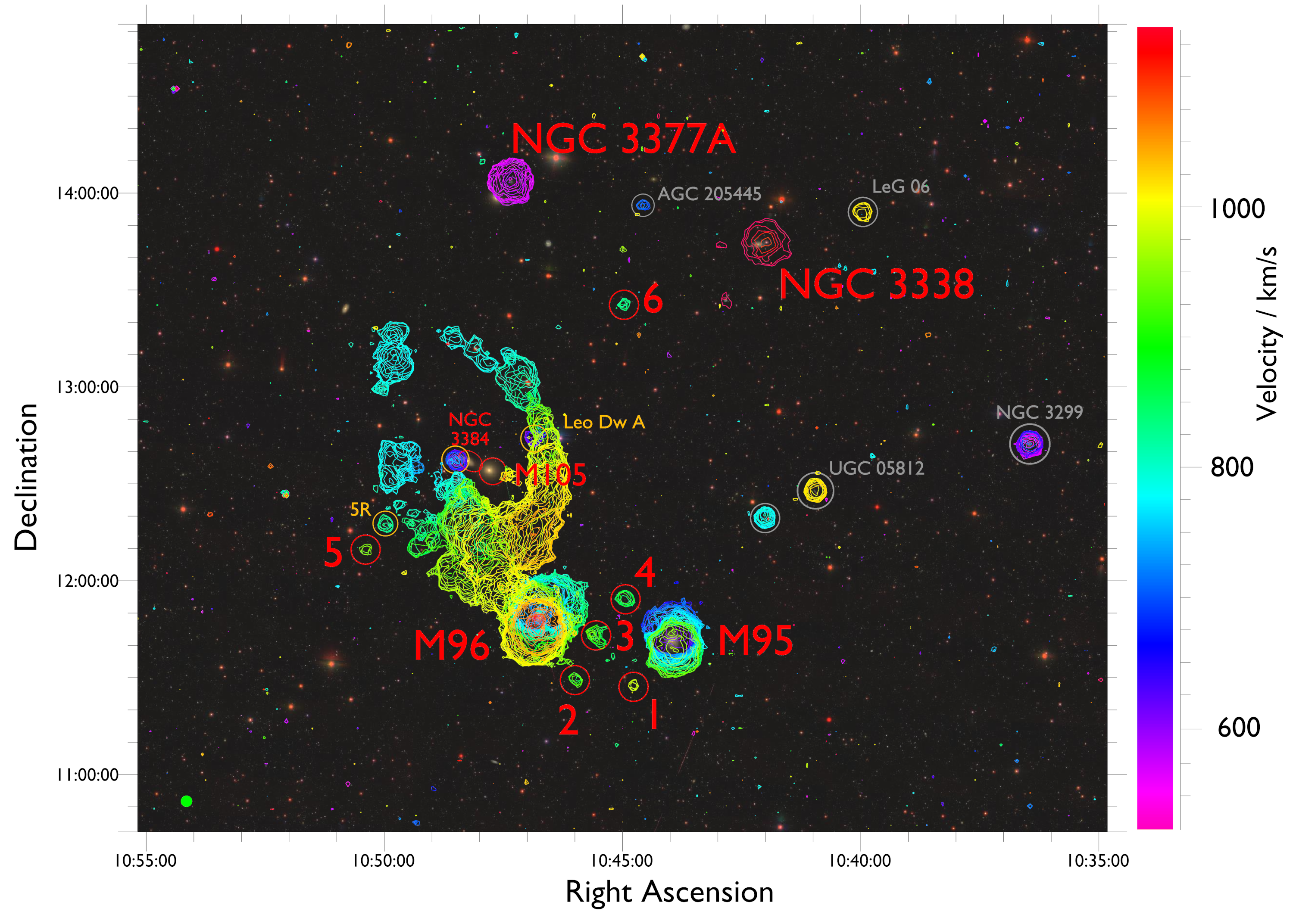}
\caption{Renzogram of the whole survey region over the velocity range of the M96 subgroup. Each \HI{} contour is at 3.5$\sigma$ and coloured according to velocity, overlaid on an RGB image from the SDSS. The Arecibo beam size (3.5$^{\prime}$, 11.3 kpc) is shown as a filled green circle in the lower left. \HI{} clouds are numbered and selected major galaxies labelled in red. Other objects used for a few comparisons but not examined in detail are highlighted in orange, while other known objects not used in this study (here shown for the sake of completeness) are highlighted in grey.}
\label{fig:wholeleo}
\end{figure*}

\HI{} spectra and SDSS images, overlaid with \HI{} contours, are shown in figures \ref{fig:cloudset1} and \ref{fig:cloudset2}. Following \cite{me20}, the contour level is chosen to be 3.5$\sigma$. We found that sources present at this level which appear to span at least one beam and three channels can be considered reliable detections, which is the case for all of our objects. Cloud 3 appears to be marginally resolved, while all the other clouds are point sources. We show a map of the full AGES Leo field in figure \ref{fig:wholeleo}. An interactive 3D version is included \href{http://www.rhysy.net/Resources/WholeLeo_Interactive_Blend4Web_v9.html}{at this url}.

\subsection{Other wavelengths}
\label{sec:otherlambda}
The presence of the giant \HI{} ring has attracted many searches for optical features of low surface brightness, including dedicated deep optical imaging in \cite{schnleo}, \cite{fergie}, \cite{leocollide} and \cite{watkins}, the latter reaching a sensitivity of $\mu_{B} = 31.0$\,mag\,arcsec$^{-2}$. In addition, \cite{kk} used reprocessed Palomar Sky Survey data to search for low surface brightness galaxies, while \cite{mull} performed sensitivity enhancement on SDSS data to search for Ultra Diffuse Galaxies. We searched the catalogues from all of these papers. In addition, we stacked both the SDSS \textit{g}, \textit{r} and \textit{i} band data and (separately) the Digitized Sky Survey blue and red images, searching the position of the clouds within the Arecibo beam area.
	
We found no hint of any optical counterpart for clouds 1, 2, 3 and 6. While there is a galaxy (LeoGroup 46) 20$''$  from the \HI{} coordinates of cloud 5, the SDSS optical redshift is $>$\,28,000\,\kms{} and it is therefore unassociated. Only cloud 4 has a clear detection at optical wavelengths : the galaxy LeG 13 (AGC 202024). We regard the optical counterpart as being almost certainly associated with the \HI{} emission even though there is no corroborating optical redshift measurement, as the optical and \HI{} coordinates differ by just 13$\arcsec$ (with no other similar objects nearby). Furthermore the optical component is sufficiently resolved that it appears more likely to be a group member than a more distant background object. There are no instances in the rest of the AGES sample of finding a single optical counterpart candidate so close to the \HI{} coordinates in which, where available, the optical redshift determination did not closely match the \HI{} measurement. The possibility of a coincidental alignment of the optical and \HI{} components can therefore be neglected.

Although it is only marginally more massive than the other clouds, cloud 4 is also the only cloud in this sample detected by ALFALFA (\citealt{aaleo}, \citealt{aa100}). The ALFALFA data (\citealt{aaleo}) gives a slightly lower \HI{} mass (6.4$\times$10$^{6}$\,\Msolar{}) than the AGES estimate (9.0$\times$10$^{6}$\,\Msolar{}), and the estimated velocity widths from ALFALFA (W50 = 24, W20 = 37 \kms{}) are also slightly lower than the AGES values (W50 = 31, W20 = 46 \kms{}). This difference is unsurprising given the differing sensitivity levels. We hereafter only use the more sensitive AGES measurements, which have a high S/N level and a detection that is clear and unambiguous in the spectrum.

The ALFALFA SDSS catalogue (\citealt{aasdss}) provides two estimates of the stellar mass for LeG 13 (i.e. cloud 4) -- we use the mean value of 6.6$\times$10$^{6}$\Msolar{}; the individual values differ by a factor three. This gives an M\HI{}/M$_{*}$ ratio of 1.4 (no star formation rate estimate is available in the catalogue as it is not detected in WISE band W4 and does not have a UV flux measurement in the NASA-Sloan Atlas, the criteria required for the star formation estimates of \citealt{aasdss}). Given this ratio, and that the \HI{} mass is only marginally higher than the other clouds, it is not obvious why cloud 4 has an easily detectable optical counterpart but the others are, if perhaps not actually dark, then significantly fainter. However, we caution that LeG 13 is itself only detected at about 6\,$\sigma$ in the SDSS. This implies that the M\HI{}/M$_{\ast}$ ratio of the other detections need not be extraordinarily high for the optical counterpart to be undetectable, perhaps only of order a few (contrast this with the more massive \HI{} detection of \cite{jozsa} which has a ratio in excess of 1,000), though the deeper optical imaging of the studies cited above would likely have revealed such an object.

\section{Analysis}
\label{sec:analysis}
There are two main possibilities for the nature of the clouds : transient tidal debris, and long-lived galaxies in which the \HI{} resides in a dark matter halo and the line width indicates rotation. We follow the analysis presented in \cite{me16} for the clouds discovered in the Virgo cluster.

\subsection{Size of the clouds}
\label{sec:size}
Cloud 3 is marginally resolved, with the 3.5$\sigma$ contours spanning approximately two beams. This gives a diameter of $\sim$20 kpc. Although the other clouds are not resolved, the large size of this cloud and its similar mass tentatively suggest that the others are unlikely to be much smaller than the beam size.

At a typical mass of 5.0$\times$10$^{6}$\,\Msolar{}, a source filling the 11.3 kpc diameter beam would have a column density of 0.05
\Msolar{}\,pc$^{-2}$ (6.3$\times$10$^{18}$\,cm$^{-2}$). Conversely, to have a column density of the more typical 6\,\Msolar{}\,pc$^{-2}$ found in dwarf galaxies (\citealt{leroy}), the diameter would be around 1 kpc. Such a column density is technically possible, as the circular-average column density may be misleading : since even cloud 3 is only marginally resolved, it is possible that it is much smaller along one axis. But, while a higher column density is consistent with observations, it is unlikely that this is actually the case. At densities comparable to dwarf galaxies we would expect to see star formation and therefore optical counterparts. The true size of most of the clouds is therefore likely of the order of a few kpc, especially given the size of cloud 3; supporting evidence of this can also be found in section \ref{sec:env}.

We can more confidently rule out that the clouds might be gravitationally self-bound by their \HI{} mass alone. As with the Virgo clouds, this would require an unphysically high column density ($>>$100\Msolar{}\,pc$^{-2}$, a value which is unknown for \HI{}, with a diameter of 0.1 kpc) even given the Leo clouds lower velocity widths. We would certainly expect to see obvious optical counterparts at densities this high, though only cloud 4 has any direct association with an optically bright galaxy. Note that while cloud 4 does have the highest \HI{} mass of the clouds, it is less than factor two greater than their median \HI{} mass.  

\subsection{Expansion time}
\label{sec:expand}
If the clouds are unstable debris, the line width can give an indication of their detectable lifetime. For the high velocity dispersion clouds in Virgo, we found that the lifetimes were so short as to make a tidal debris hypothesis unlikely. In contrast the Leo clouds have substantially narrower line widths, which does not give such a stringent constraint on their evolution. For cloud 3, which is marginally resolved, to reach its 11 kpc radius at its presumed expansion velocity (W50/2 = 21 \kms{}) would take 0.5 Gyr. Therefore, with similar expansion speeds and assuming sizes half that of cloud 3, the other clouds could have existed for $\sim$250 Myr. 

Although obviously approximate, these estimates can be considered upper limits of the cloud survival times based on their line widths. We have assumed the clouds began as point sources, which is unphysical - a more realistic initial size could only lead to shorter expansion times. Likewise we assume that the clouds fill the beam, whereas most of them are probably somewhat smaller (but see section \ref{sec:env} for an important caveat). More significantly, in some cases the W20, which is much larger than the W50 value, is a more accurate estimate of the true line width (see section \ref{sec:btfr}), and using this value for the expansion velocity would roughly halve the timescale estimates. Even so, the long lifetime estimates are at least consistent with the clouds being tidal debris, but we consider this further in section \ref{sec:env}.

\subsection{Dynamical masses}
The optical counterpart of cloud 4, AGC 202024 (LeG 13), allows for different interpretations of the nature of the clouds. One possibility is that the clouds are tidal debris in different stages of evolution, with only cloud 4 as yet having undergone any significant star formation or perhaps the only one in which star formation continues (see \citealt{fadug} for a discussion on fading tidal dwarfs). Another option is that most clouds are tidal debris but cloud 4 is an ordinary galaxy, and its close position and similar \HI{} properties to the other clouds is purely coincidental. It is also possible that the clouds are in fact \textit{all} primordial galaxies (e.g. satellites), with their \HI{} embedded in dark matter halos. In this case cloud 4 would be significantly optically brighter than the others, though this does not necessarily mean the other clouds are totally dark. In this interpretation, we can use their line widths to estimate their dynamical masses, i.e. :

\[M_{dyn} = \frac{r\,v_{circ}^{2}}{G} \]

For cloud 3 we may take the radius to be 11 kpc since it is marginally resolved. With a rotation speed $v_{circ}$ of W50/2 = 21 \kms{} (note that though the line width is small, it does appear to show a velocity gradient - though we lack sufficient resolution to say if this is the result of ordered motions), this gives M$_{dyn}\geq$1$\times$10$^{9}$\Msolar{} (this is a lower limit as we do not correct for inclination), giving a ratio M$_{dyn}$/M\HI{}$\geq\,$140. This would be an extremely dark matter-dominated object.

For the other clouds, assuming a radius of 5.5 kpc (half the beam size) and a typical line width 30 \kms{}, dynamical masses would still exceed 3$\times$10$^{8}$\Msolar{}, with M$_{dyn}$/M\HI{}$\,\gtrsim\,$50. At the smallest plausible size of a radius of 0.5 kpc (see section \ref{sec:size}), M$_{dyn}\,\geq$3$\times$10$^{7}$\Msolar{} and M$_{dyn}$/M\HI{}$\,\gtrsim\,$5. Despite their low line widths, the clouds are consistent with requiring a significant quantity of dark matter in order to be stable and long-lived.

\subsection{Environment of the clouds}
\label{sec:env}
\subsubsection{Clouds 1-4}
\begin{figure}
\centering
\includegraphics[width=80mm]{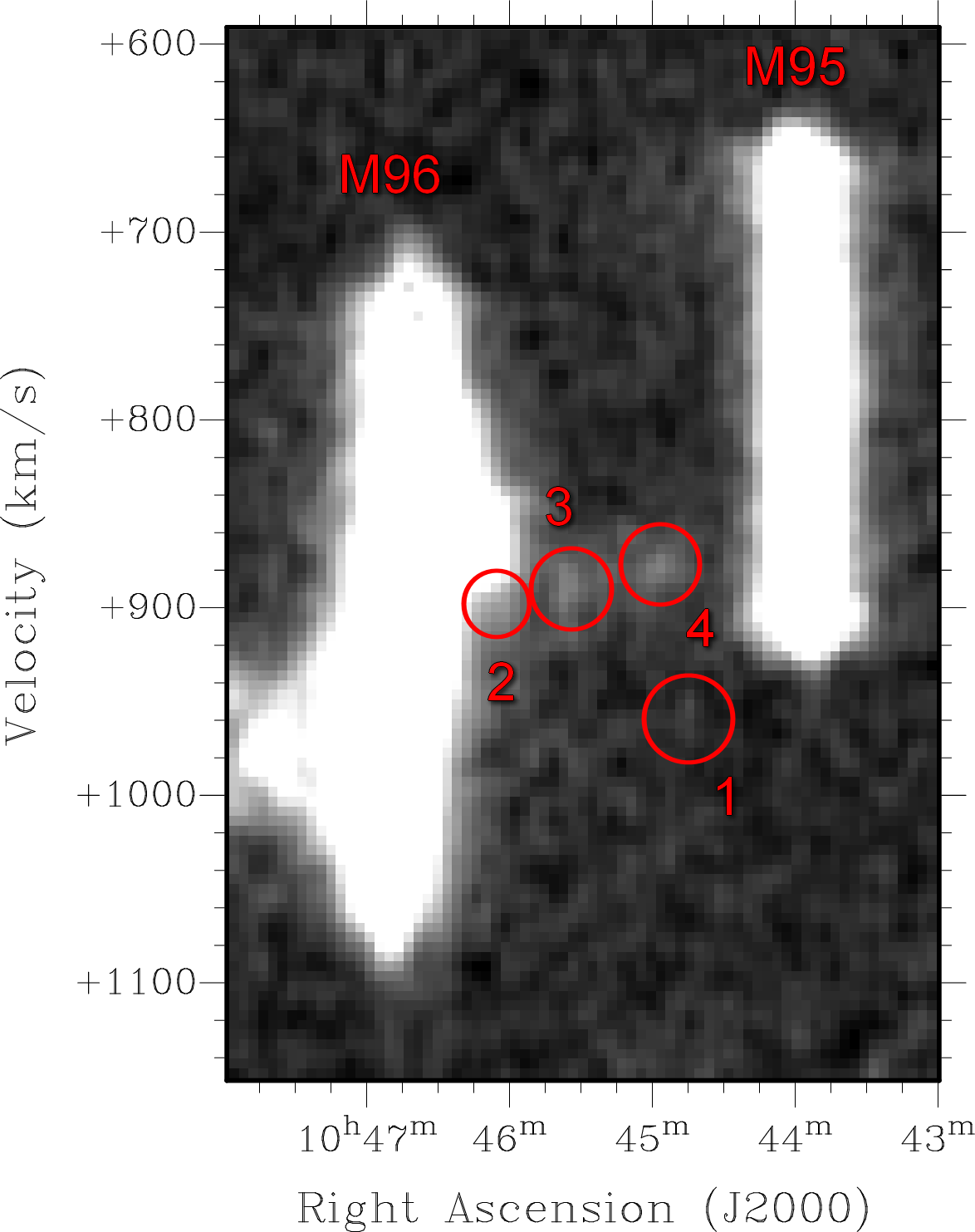}
\caption{PV diagram of the M95-M96 region with a linear colour stretch, highlighting clouds 1-4.}
\label{fig:pvclouds}
\end{figure}

The proximity of the nearest galaxies is crucial to understanding the origin of the clouds. Clouds 1-4 are midway between the giant spiral galaxies M95 and M96, which have a projected separation of approximately 0.7 degrees, equivalent to 136 kpc. The clouds are also at similar velocities to the two spirals (M95 is at 779 \kms{}, M96 is at 888 \kms{}), as shown by the PV-diagram in figure \ref{fig:pvclouds}. Cloud 1 is however somewhat deviant from the trend in P-V space seen from clouds 1-4.

In section \ref{sec:expand} we calculated that the clouds could have survived for as long as 250 Myr even if in free expansion. To travel the approximate 70 kpc from the nearest spiral in 250 Myr would require a projected velocity of 270\,\kms{}. In comparison, the  dispersion of the Leo I group as a whole is 175\,\kms{} (\citealt{aaleo}), with the line-of-sight velocity differences of the clouds from M95/M96 ranging from 6 to 181\,\kms{}. However 70 kpc is considered from the distances to the centres of the galaxies. A more realistic value, accounting for the apparent size of the \HI{} discs, would be about 40 kpc, reducing the velocity to about 150\,\kms{}. Given that the survival timescales are approximate upper limits, and that the true 3D velocity must necessarily be higher than the projected velocity estimate, it seems reasonable to say that the clouds are consistent with being tidal debris, albeit with velocities towards the higher end expected for objects in this environment.  

Perhaps more problematic for this interpretation is that neither M96 nor M95 show much indication of a significant disturbance in their \HI{} discs. M95 in particular shows no signs of abnormality - by the criteria established in \cite{me20}, we would identify this galaxy as undisturbed. In contrast, M96 does show evidence of a warp (this is much easier to see in the online 3D tool than the renzogram or PV map), as well as being asymmetrical in its optical appearance, but not any larger extension that could be identified as a tail that would indicate gas removal. 

\begin{figure*}
\centering
\includegraphics[width=180mm]{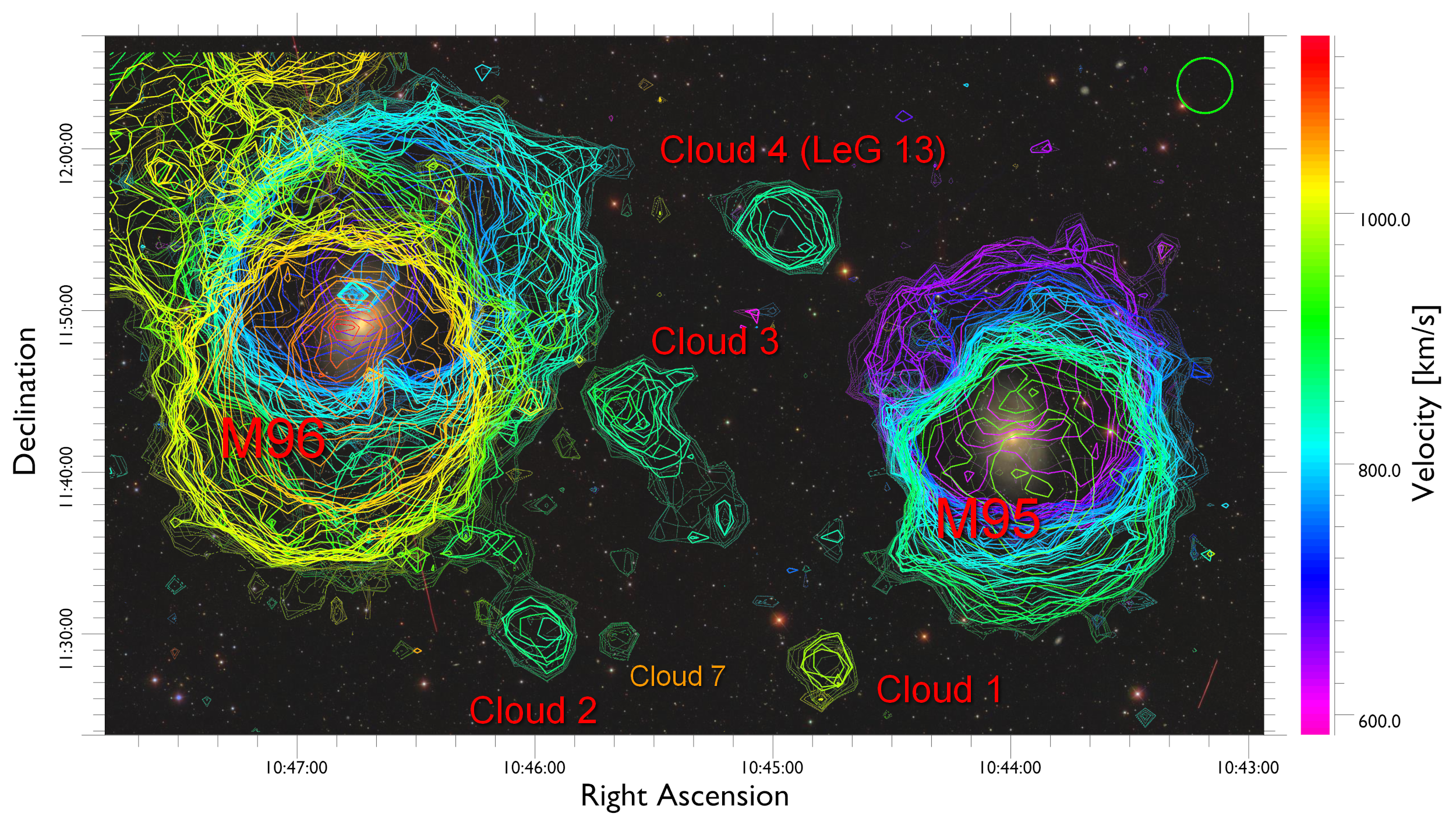}
\caption{Renzogram of the M95-M96 region. The thick contours are from the standard AGES cube at 3.5$\sigma$. The thin contours are from a cube with additional Hanning smoothing (width 15) at 4$\sigma$. The green circle in the upper right shows the AGES beam. An additional feature tentatively identified as cloud 7 is also shown.}
\label{fig:m95renzo}
\end{figure*}

One possibility is that the clouds in this region are the brightest peaks in a lower-density bridge of material connecting the two galaxies, as suggested for the M31-M33 clouds (\citealt{wolfe}). If so, the large-scale \HI{} feature expected in tidal encounters (\citealt{me16}, \citealt{me17}) would exist but happen to be below our column density sensitivity. We investigate this in figure \ref{fig:m95renzo} (see figure \ref{fig:wholeleo} for a map of the whole survey region). The figure shows renzogram contours at the 3.5$\sigma$ level from the standard AGES cube, and also thinner contours at 4$\sigma$ from the cube after additional Hanning smoothing (width 15 channels, equivalent to 40\,\kms{} resolution) of the spectral axis - spectral smoothing has the advantage of increasing sensitivity without degrading spatial resolution. The smoothing reduces the $rms$ to 0.3 mJy, about half that of the standard cube. Note that due to the lower spectral resolution, the sensitivity to \textit{total} mass present, i.e. column density, is degraded to N$_{\HI{}}$ = 3.0$\times$10$^{17}$cm$^{-2}$ at 1$\sigma$ - smoothing improves the sensitivity to the \textit{average} mass present per unit velocity, which is the relevant factor when searching for diffuse emission.

Interestingly, this extra smoothing does not alter the appearance of M96, M95 or cloud 1, but the other clouds do appear more extended. Cloud 2 shows a tentative connection to M96. Somewhat surprisingly, cloud 4 appears to be marginally resolved, spanning approximately two beams - given that it has an optical counterpart, we would expect it to have a higher column density than the other features and so be more compact, yet this is not the case. However, this is a further indication that our size estimates in section \ref{sec:size} are unlikely to be wide of the mark. Cloud 3 becomes significantly more extended, spanning four beams (45 kpc) with a clear velocity gradient across the whole feature (the velocity difference between the most extreme contours where the cloud is visible is 67\,\kms{}), though it does not appear to directly connect with the disc of M96.

The extra Hanning smoothing also reveals a tentative additional detection here labelled as cloud 7. In the standard cube this is just visible but appears as an extension of cloud 2. Owing to its small size (the contours do not even span a whole Arecibo beam) and low S/N (peak 5.4, integrated 6.25), we treat this detection with caution. If real, it would have an \HI{} mass of 3.5$\times$10$^{6}$\,\Msolar{}, a W50 of 52\,\kms{} and a W20 value of 80\,\kms{}.

Despite the detection of extended features and a possible additional cloud, the extra processing does not reveal any large-scale bridge of material linking M95 and M96 - at this sensitivity level, there is no evidence that the clouds are peaks in a larger structure. While cloud 2 does resemble some of the clouds discussed in \cite{olivia} (where features which appeared as discrete at a lower sensitivity were revealed as being connected to the disc of M33 at a higher sensitivity),  the smoothing does not suggest the disc itself is likely to be much larger than it appears in the standard cube. This does not rule out a tidal origin for the clouds, but does not fit the general expectation of tidal stripping.

Of course, the elephant in the room is the Leo Ring - M96 is directly connected to the main body of the Ring\footnote{The impression one gets from inspecting the data is that the \HI{} of M96 is superimposed on that from the Ring. This is somewhat subjective, but clearly there is nothing like the classical tail and counter-tail structure predicted in most tidal encounters (\citealt{tooms}).}, and it is possible M95 was also involved in the creation of this feature. Potentially M96 could be interacting with the other major galaxies in the Ring (e.g. M105, NGC 3384) as well as M95. It is interesting to note that the clouds 2, 3 and 4 are close in velocity to the warp seen in the \HI{} of M96 -- again, in PV space these clouds do resemble a bridge, but this is not seen in PP space.

Finally, another scenario is that the clouds are not the result of a direct interaction between M95 and M96 but rather result from whatever process formed the Leo Ring. The appeal of this explanation is that the Leo Ring has a velocity width of $\sim$400\,\kms{}, making it much easier for the clouds to reach a large separation from their parent galaxies (see also section \ref{sec:clouds5-6}). On the other hand the kinematics of the clouds do not match the velocity gradient of the Ring at all, being at velocities approximately 150\,\kms{} lower than the material in the nearest point on the Ring.

Though these caveats are important, overall the evidence would still seem to point towards a tidal interpretation for clouds 1\,--\,4. They lie midway between two galaxies, one of which shows signs of disturbances (albeit more weakly than we would expect), and at a distance consistent with the clouds being unstable debris. The lack of a large stream of \HI{} is, in our view, a relatively minor difficulty in this case. Despite the increase in sensitivity from smoothing, it is still possible that such a stream exists but is below our detection threshold, with the clouds being significant density enhancements within it (as in. \citealt{wolfe}). In our Virgo simulations (\citealt{me16}, \citealt{me17}), we found that clouds themselves detached from their parent galaxies were usually found near large, easily detectable \HI{} tails which were directly connected to their parent galaxies. Such features are not seen here unless we count the Ring itself, but the cluster and group dynamics are markedly different. Simulations of low velocity dispersion groups could help address whether (and for how long) we should expect large streams to accompany detached clouds in these environments, and/or whether the detached clouds themselves are an expected result of tidal encounters.

\subsection{Clouds 5 and 6}
\label{sec:clouds5-6}
While it does appear distinct, cloud 5 is in close proximity to the Leo Ring. This region of the Ring is highly disorganised, and it is possible this is an outlying cloud formed by the same processes that produced the rest of the large-scale \HI{} emission in this region. To the north, a much larger and more massive cloud is also detached but appears to be part of the Ring. Whereas the northern feature clearly fits neatly into the ellipsoid structure of the rest of the \HI{}, cloud 5 is not aligned with any other features. Moreover, the additional smoothing described for clouds 1\,--\,4 does not reveal any additional component to cloud 5, and its narrower line width suggests that it is - at least at present - truly detached from the Ring and not an outlying part of the larger object. Still, the most obvious explanation is that cloud 5 was produced along with the rest of the Ring, a process we will explore in a future work.

Cloud 6 is much more intriguing. While it is possible that it too was formed by the same mechanism that created the ring (e.g. the collisional origin proposed by \citealt{leocollide}), its distances and kinematics argue against this interpretation. It is 0.65 degrees away from the nearest point on the Ring, about 125 kpc in projection, and does not align with any features in the Ring. The distance to M105, at the approximate centre of the Ring, is about 1.1 degrees, 210 kpc. At the estimated evolution time of 250 Myr (section \ref{sec:expand}), this would require a projected velocity of 820\,\kms{}. Almost all of this would have to be across the plane of the sky, since the velocity of the cloud falls well within the velocity of material of the Ring. In contrast, the Ring itself only spans about 400 \kms. The kinematics of the cloud and the Ring thus do not match : it is very difficult to see what sort of process could produce a gigantic, coherent arc of material and also one single, compact cloud that was ejected at least at twice the velocity of everything else. While we cannot absolutely exclude a common origin of the Ring and cloud 6, there are other possible candidate parents for the cloud. 

There are four other nearby \HI{}-detected galaxies that suggest themselves as potential parents based on their angular separation from cloud 6. Two of these, UGC 05832 (a disturbed spiral) and CGCG 065-090 (an irregular) form a close pair, and we cannot fully distinguish the \HI{} detection of the individual galaxies in this case. These are at least 0.47 degrees away in projection (90 kpc) from the cloud, with a velocity separation of 360 \kms{}. As discussed in the introduction, this velocity difference is much greater than the group velocity dispersion, making an association less likely. Near to this pair of galaxies is a third, much more massive spiral NGC 3338 (note that while the pair are not visible in figure \ref{fig:wholeleo} as they are outside the velocity range shown, part of NGC 3338 is just visible). This is rather further away at 0.77 degrees (147 kpc) and even more separated in velocity (440\,\kms{}) from the cloud. While the spiral/irregular pair do show disturbances in their optical morphology, none of these three galaxies show significantly extended \HI{} emission : we can easily distinguish the \HI{} in the pair from the giant spiral, and there is no hint of any extension towards cloud 6. 

In addition, NED gives a mean redshift-independent distance estimate to NGC 3338 of 22.6 Mpc, twice the distance to the M96 subgroup. Likewise, the fact that \textit{all} of the clouds are found within a velocity range of 100 \kms{} while UGC 05832 and CGC 065-090 are found at a difference of 360 \kms{} argues against their association with any of the clouds. Again, given the low velocity dispersion of the Leo group, it seems much more probable that all the clouds are found at this much lower distance and are unrelated to any of the background galaxies.

The fourth candidate is the spiral NGC 3377A, rather uniform in morphology, which is 0.86 degrees (164 kpc) away in projection and separated in velocity by 287 \kms{} from the cloud. As with the other candidates, this galaxy also shows no hint of any \HI{} extension and its velocity separation is quite high compared to the dispersion of the M96 subgroup.

Frustratingly, none of these candidates can be definitively assigned or eliminated as the parent of cloud 6. The \HI{} masses of all four candidates (\citealt{aaleo}) range from two to four orders of magnitude greater than the clouds, implying that we should easily be able to detect any disturbance in the parent galaxy sufficient to account for the clouds. The low mass of the clouds means that estimating the \HI{} deficiencies of the candidate parent galaxies would be useless since they would only constitute a small fraction of any missing gas (see \citealt{me17} for a discussion on the well-known, more massive example of VIRGOHI21 and its parent galaxy NGC 4254).

\subsection{Comparisons with other clouds}
\label{sec:others}
In \cite{me16} we presented a table of isolated dark clouds with possible primordial origins. Very few clouds are known which are similar to the Leo clouds. We briefly review those with notably similar features below. These were selected on the basis of being clearly detached from their parent galaxies (which are usually difficult to identify) and having no large-scale extended \HI{} emission (such as tails) visible in their vicinity.

\subsubsection{GBT1355+5439}
Discovered in \citealt{mihosm101} and with follow-up observations described in \citealt{oo13}, this cloud is located close to M101 in projection. Due to its low systemic velocity (210 \kms{}), the distance is highly ambiguous. If this object is in the Local Group, then \cite{oo13} propose it could be a dark matter minihalo, with a size of about 1 kpc and M\HI{} $\sim$1$\times$10$^{5}$\,\Msolar{}. At the distance of M101, it would be 150 kpc from M101 with an \HI{} mass just a bit larger than the Leo clouds at 1$\times$10$^{7}$\Msolar{}. The W20 is given in \citealt{mihosm101} as 41 \kms{}; W50 is not reported. \citealt{oo13} concluded that there is no clear explanation for the object, with all of the proposed interpretations (galactic mini-halo, dark galaxy, tidal debris) having advantages and disadvantages.

\subsubsection{GEMS\_N3783\_2}
Discovered in \cite{kilborn} in a survey of the NGC 3783 group, this cloud is considerably more massive than the Leo clouds at 3.8$\times$10$^{8}$\Msolar{}. The W50 is 106 \kms{} and the W20 116 \kms{}. Although the nearest spiral does show an extended, distorted \HI{} disc that might indicate an interaction, it is 450 kpc away from the \HI{} cloud and does not show any long \HI{} tails. \cite{kilborn} do not rule out a dark galaxy interpretation, but consider a tidal origin more probable.

\subsubsection{NGC 1395 clouds}
\cite{wong} discovered two clouds (C1 and C2) in the vicinity of NGC 1395, of mass 2\,--\,3$\times$10$^{8}$\Msolar{} - again considerably more massive than the Leo clouds. These clouds appear to be of similar velocity width to the Leo clouds, with W50 measured at 17\,--\,41 \kms{} from ATCA and Parkes respectively (W20 is not reported - from the figures, W20 may be substantially higher but affected by noise). Unfortunately one cloud is projected in front of the elliptical galaxy NGC 1403, hampering the identification of any optical counterparts (NGC 1403 itself almost certainly has no association with the cloud, it is at a systemic velocity more than 2,000 \kms{} higher than the \HI{} detection). The clouds range from 240\,--\,360 kpc in projected distance from NGC 1395. In marked contrast to the Leo clouds, \cite{wong} note that the line width and size of these clouds are consistent with being gravitationally self-bound by the mass of the \HI{} alone. Their favoured hypothesis is that the clouds could be optically dim or dark tidal dwarf galaxies, though the lack of any observed tidal tails in the region is arguably problematic for this interpretation.

\subsubsection{SECCO 1}
SECCO 1, also known as AGC 226067, is discussed in \cite{adamsclouds}, \cite{bellazz} and \cite{secco}. The object is an optically faint (but not dark) \HI{} cloud in the direction of the Virgo cluster. Candidate parent galaxies are at least 250 kpc away in projection and the stellar component appears to lack an old population. Its \HI{} mass is a bit larger than the Leo clouds at 1.5$\times$10$^{7}$\Msolar{} (at the Virgo cluster distance), as is its W50 line width at 54 \kms{}. Simulations in \cite{bellazz} show that a low line width cloud could survive for over 1 Gyr moving through the intracluster medium, certainly long enough for the cloud to have reached its separation from its candidate parent galaxies. What remains unclear is why, if there is really no old stellar population, the star formation should apparently have begun only very recently despite no obvious source of perturbation.

\subsection{The baryonic Tully-Fisher relation}
\label{sec:btfr}
\begin{figure*}
\centering
\includegraphics[width=160mm]{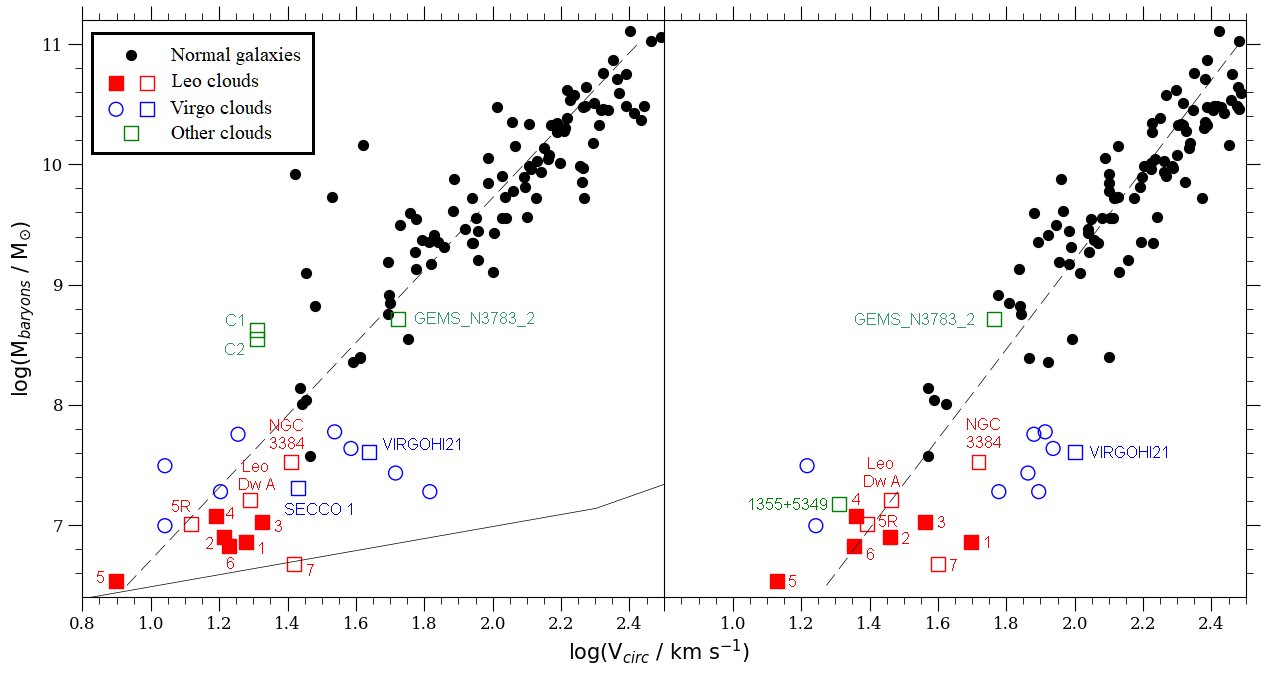}
\caption{Baryonic Tully-Fisher relation for normal galaxies (black points, from the AGES Virgo background fields) and a selection of optically dark \HI{} clouds, using the W50 (left) and W20 (right) estimates for the line width. The line widths are corrected for inclination for the galaxies but not the \HI{} clouds. The baryonic mass for the clouds is computed using their \HI{} mass multiplied by a factor 1.36 to account for the presence of helium. The dashed line is the best fit to the optically bright galaxies. Our main sample of Leo clouds are shown with red squares while additional objects in Leo are shown as open red squares (highlighted with orange circles in figure \ref{fig:wholeleo}; cloud 7 is shown in figure \ref{fig:m95renzo}). Clouds from the Virgo surveys of AGES are shown with open blue circles while other clouds in the Virgo cluster are shown as labelled open blue squares. Other clouds in green are described in section \ref{sec:others}. The solid line in the left panel shows the completeness line for an integrated S/N ratio of 6.5.}
\label{fig:btfr}
\end{figure*}

One of the most intriguing features of the AGES Virgo clouds is their offset from the baryonic Tully-Fisher relation. Most normal galaxies appear to show a tight correlation between their rotation velocity and baryonic mass (\citealt{btfr}). While some ultra-diffuse galaxies have been found to have velocity widths well below the expectations from the BTFR, given their baryonic mass (\citealt{pina}\footnote{While not directly comparable in terms of the BTFR, systems with a similar apparent deficit of dark matter are discussed in, for example, \cite{vd} and \citealt{guo}.}), the AGES Virgo clouds of high line widths show the opposite behaviour. In figure \ref{fig:btfr} we compare the AGES Virgo and Leo clouds on the BTFR, along with a selection of other optically dark clouds from other surveys. We also include other features found in Leo : the tentative feature dubbed cloud 7 (see section \ref{sec:env}), the faint galaxy Leo Dw A (figure \ref{fig:wholeleo}, \citealt{schnleo}), the cloud adjacent to NGC 3384 (figure \ref{fig:wholeleo}, note that the measurements are uncertain due to the close proximity of material from the Ring), and a `cloud' labelled 5R (again see figure \ref{fig:wholeleo}) which is likely part of the Ring itself.

The results vary depending on whether we use the W50 or W20 estimators for the line width. Before discussing this in relation to the clouds, it is important to note that this is also true for normal galaxies, with three optically bright galaxies having apparently anomalously low circular velocities when using W50 but normal values (i.e. lying well within the general scatter) when using W20. This is not because they are similar to the \cite{pina} objects, but only because of the profile shape : a high S/N but narrow spike can lead to an erroneously low value for W50, even when the W20 estimator is still itself at a high S/N level. We consider the W20 estimate to be generally more reliable for the Virgo clouds, and the deviation towards high velocities is seen even using W50 for some of them. It is important to note that neither W50 nor W20 are infallible measures of the true width of the line. While W20 may give overestimates due to being measured at a lower S/N level, W50 can give underestimates despite being measured at a higher S/N level (possible improvements to measuring the velocity width are described in \citealt{yu20} and \citealt{yu22}).

Some of the Leo clouds appear to follow the BTFR for bright galaxies. Ostensibly, there is no particular reason to expect tidal debris to follow the BTFR - non-rotating, unbound debris (regardless of its formation mechanism, tidal or otherwise) has no constraints to follow the same relation as for rotating, stable galaxies. In the case of the Leo clouds, most seem likely to have formed as a result of interactions, based on their proximity to galaxies and/or the Leo Ring, with the notable exception of cloud 6. Yet most of the clouds in our Leo sample do seem to follow (or are at found close to) the BTFR for normal galaxies -- including features which seem to be part of the Ring (the single exception is the tentative detection of cloud 7, which we discuss further below). An important caveat is that, using W50, most of the clouds are found at higher velocity widths than expected, with only cloud 5 at lower velocities (though clouds 4 and 5R are very close) -- if the clouds were really following the relation, we might have expected about half to have higher and half to have lower widths than the general relation.

There is no obvious reason why the clouds should follow (or be close to) the BTFR for bright galaxies, as it cannot be a selection effect. Given the typical maximum column density of \HI{} of $\sim$10\,\Msolar\,pc$^{-2}$ (\citealt{leroy}) and the beam size, we could expect to detect up to 10$^{9}$\,\Msolar{} of \HI{} within the Arecibo beam at the 11.1 Mpc of the group -- about two orders of magnitude greater than the BTFR implies for clouds of these line widths. Indeed, as discussed, dark clouds are known elsewhere which are considerably more massive than the Leo clouds and which do show a strong deviation from the BTFR. 

Conversely, we can also consider how large the line widths of the clouds would be for them to become undetectable if their mass was kept constant. Although the clouds have narrow line widths, they are reasonably bright. By the prescription of \cite{aasn} we should be complete to sources of these typical masses (equivalently, $\sim$\,0.2 Jy\,\kms{} flux) for velocity widths of up to 130 km/s, far wider than their actual values. In short, there is nothing prohibiting the existence of much more massive clouds given their observed $\sim$30\,km/s widths, nor any reason they could not be detected at much higher line widths for the same total flux : they could be detected over a much larger part of the BTFR parameter space than the region in which they are actually found.

Examining the W20 estimator complicates the situation. Using W50, most Leo clouds are found at somewhat higher velocities than the fit for optically bright galaxies, even if only slightly. For the W20, most lie closer to the fit, but there are some stronger deviations as well. With W50, of our Leo objects only cloud 7 shows an appreciable deviation from the standard BTFR, whereas with W20, a similar deviation is also seen for cloud 1 and possibly the cloud adjacent to NGC 3384. The faint nature of cloud 7 implies a need for observations of high sensitivity and especially of higher spatial resolution than AGES to determine if this is really a discrete object or only part of cloud 2, and similarly accurate measurements of the NGC 3384 cloud are confused by intervening material from the ring -- it is likely that the W20 measurement is an overestimate in this case. Intriguingly, the W20 measurement of cloud 1 appears to be accurate - indeed, if anything it seems more likely that the W50 is an underestimate. We note that, unlike the other clouds, its spectral profile is clearly asymmetric (figure \ref{fig:cloudset1}), suggesting that it might have multiple components and reinforcing the need for higher resolution observations. Based on their proximity to other galaxies and \HI{} structures, all of the Leo objects with excessively high W20 values are likely tidal in origin.

An important caveat is that the results discussed above are sensitive to the best fit line used for the BTFR, which in figure \ref{fig:btfr} is heavily dominated by optically bright galaxies which are more than two orders of magnitude brighter than the optically dark clouds. For this figure we used the same basic methodology as in \cite{me13} : the baryonic mass is the combination of the measured M\HI{} (with an assumed correction for helium) plus the stellar mass, with the line width corrected only for inclination. This approach has the advantage of simplicity and relies entirely on our own measurements. A more sophisticated approach necessitates greater complexity in correcting the measurements, but allows us to compare our results with the gas-dominated, lower-mass sample presented in \cite{btfr12}. This should provide a more robust comparison with the Leo clouds, in particular a quantitative assessment of how well they agree or deviate from the BTFR for normal galaxies given the general scatter. Importantly, \cite{btfr12} used resolved rotation curves to estimate the circular velocity which gives much more accurate results than line widths. Note that the corrections applied to our own data are extensive and fully described in the appendix. The result is shown in figure \ref{fig:btfrmcg}.

\begin{figure}
	\centering
	\includegraphics[width=85mm]{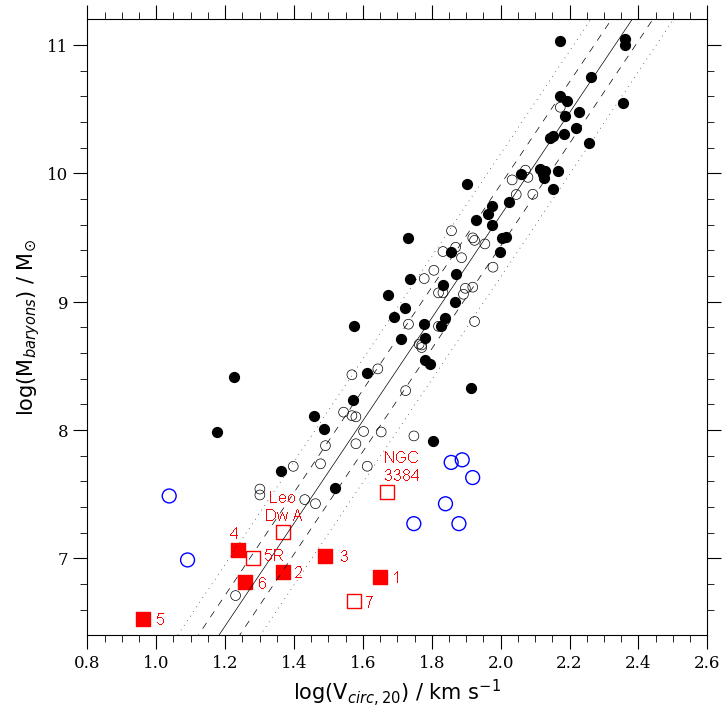}
	\caption{Baryonic Tully-Fisher relation using more sophisticated corrections for velocity width (derived for our sample from W20) and stellar mass, following the prescriptions of \cite{btfr12} and \cite{spring} -- for full details see the appendix. The colour scheme is as for figure \ref{fig:btfr}, except that the open circles show the sample of \cite{btfr12} (his table 1). The black solid line shows the best fit relation according to \cite{btfr12}, with the dashed and dotted lines respectively showing the 1$\sigma$ (0.24 dex in mass) and 2$\sigma$ scatter, again according to \cite{btfr12}. Clouds from non-AGES samples are not included as we lack the data need to make the appropriate corrections to velocity width, which is corrected for inclination, spectral resolution, and cosmological expansion.}
	\label{fig:btfrmcg}
\end{figure}

Figure \ref{fig:btfrmcg} supports the basic result of the right-hand panel of figure \ref{fig:btfr}. Clouds 1, 7 and the NGC 3384 feature clearly deviate from the BTFR in the sense of having an unexpectedly high velocity width, while cloud 5 may have a width somewhat lower than expectations. Clouds 2, 4, 6, as well as 5R and Leo Dw A, all lie well within the general scatter. Cloud 3 is a marginal case, having a velocity width slightly exceeding the 2$\sigma$ scatter. Overall, the main result appears robust : some clouds do deviate from the BTFR, but others do not. We note that there is a significant caveat in that the exact form of the BTFR depends strongly on the corrections used, especially for the stellar mass, with \cite{manybtfr, btfr12} describing how the slope -- the exponent in the power-law fit --can vary from 3\,--\,4 -- readers are strongly encouraged to consult the appendix for details.

While these deviations from the BTFR might seem to indicate that these clouds do indeed have a non-galaxian nature (i.e. they are transient and unstable), several difficulties arise. This does not explain why the other objects stubbornly remain within the observed scatter of the BTFR, including cloud 4 and arguably cloud 3, which are both close in projection to clouds 1 and 2 (with significant caveats : cloud 3's extended nature may mean its velocity width has been underestimated, while cloud 4 has an optical counterpart so may be an ordinary galaxy). Moreover, the similarly excessive velocity widths of the Virgo clouds has proved difficult to explain : in \cite{me16}, we found that it was very difficult to reproduce the high velocity widths of the Virgo clouds if they were produced as tidal debris\footnote{Essentially, the higher the line width, the faster the dispersion and so the faster the clouds reach an undetectably low S/N ratio - the more excessive the velocity widths, the shorter the lifetime of any such objects. See \cite{me17} for a full discussion.}, whereas if they were faint or dark galaxies their deviation from the BTFR could be easily explained and was robust to tidal interactions with other galaxies. However, as discussed, the dynamics of tidal encounters in a cluster will be significantly different than in a group environment, again indicating a strong need to simulate a more Leo-like region before any firm conclusions can be drawn.

Of the non-AGES clouds, even more caution is needed since the values for W50 and W20 are not reported homogeneously (they are shown only in figure \ref{fig:btfr} but not figure \ref{fig:btfrmcg} as we lack the data for corrections necessary to estimate their velocity widths). Both of the \cite{wong} clouds appear to have lower than expected velocity widths given their gas mass, even using the line widths without any additional corrections : \cite{wong} note that while the clouds do appear to be rotating, W50 likely overestimates the rotation velocity with a significant component of the line width originating from velocity dispersion. The other two clouds (GBT1355+5439 and GEMS\_N3783\_2), though differing in mass by almost two orders of magnitude, both lie on the same BTFR as for normal galaxies. The amount of intrinsic scatter on the BTFR is controversial, with \cite{pina} arguing that some galaxies are strongly deviant (see also \citealt{pina22}) whereas \cite{lelli} say the intrinsic scatter is below Standard Model expectations, and \cite{btfr12} find there is no scatter unexplained by observational errors. If the latter is correct, then it seems remarkable that this is even true for some objects which are likely debris where the usual dynamical relations should not apply. Paradoxically, it seems that some objects for which the evidence generally indicates a tidal origin (e.g. clouds 1\,--\,5 in Leo) actually lie on the standard BTFR, while some for which the evidence is - arguably - against a tidal origin (e.g. the AGES Virgo clouds, the \citealt{pina} UDGs) clearly deviate from it. This is exactly opposite to what we might expect.

\section{Summary and discussion}
\label{sec:findings}
We reported the discovery of at least five optically undetected \HI{} clouds in the M96 subgroup. One of the objects (cloud 5) is close to the giant Leo Ring, and most likely part of that structure -- we leave analysis of this to a future work. Three of the optically undetected \HI{} clouds (as well as an additional more tentative detection not included in our main sample) lie between M96 and M95, suggesting a tidal origin. In position-velocity space these clouds appear to be part of a bridge connecting the two spiral galaxies, but in position-position space this structure is not seen. No elongated structures directly attached to M95 are visible, though M96 is connected to the Ring. Very near to these three clouds, and at a similar velocity, the dwarf galaxy LeG 13 (cloud 4 in our catalogue) is detected with similar \HI{} properties to the dark clouds but with a clear optical counterpart. Elsewhere in the group, another object (cloud 6) is seen with no apparent association to the Ring or any of the galaxies present. None of its nearest \HI{}-detected neighbours show any signs of disturbances in their gas discs. With the nearby galaxies having \HI{} masses two to four orders of magnitude greater than the clouds, which are themselves readily detected, even a small perturbation to their \HI{} ought to be easily detectable with AGES -- and their distances are anyway well outside the Leo Group. This makes the parent galaxy of cloud 6 extremely challenging to identify.

All of the clouds (including LeG 13) possess comparable \HI{} masses and velocity widths, but in other aspects the clouds are dissimilar. While clouds 1-4 are all between M95 and M96, cloud 1 is offset in velocity from the others and has a higher velocity width. Cloud 2 may be connected to the disc of M96, while cloud 3 appears significantly extended when spectral smoothing is applied. Only cloud 4, LeG 13, shows an optical counterpart. Cloud 5 is near the Leo Ring, while cloud 6 is isolated.

Given this, it seems probable that the clouds formed from a variety of mechanisms, though a full exploration of this will require detailed numerical simulations. Overall, clouds 1\,--\,4 seem most likely to be tidal, but it is not at all obvious why cloud 4 alone is optically bright while its neighbours are undetected (it is tempting to suggest that as it has the highest \HI{} mass it might have the highest column density, but its mass is only marginally greater than the others and it shows some hints of being more extended as well). Cloud 6 is least likely to be tidal given its separation from its nearest galaxies, but the low line width of all clouds would allow them to survive travel across distances exceeding 100 kpc at velocities compatible with the group velocity dispersion. The main difficulty for the tidal debris interpretation is the lack of any elongated tails found around any of the galaxies in this region, the Leo Ring itself notwithstanding. If the clouds were produced by the same process that created the Ring, then they provide additional constraints for any future modelling efforts -- in particular, it is not obvious how cloud 6 could be explained by a simple galaxy-galaxy collision model.

A further oddity for the tidal debris scenario is that most clouds seem to obey the baryonic Tully-Fisher relation seen for normal, stable galaxies. This is unexpected as the clouds are bright enough that they could be detected at much higher line widths (with the caveat that they would have shorter detectable lifetimes) and large enough that they could have much higher gas densities before star formation would be expected. In essence, both their baryonic masses and velocity widths are not constrained by selection effects to obey the BTFR, but this is where the clouds are found nonetheless. 

Ironically, it may be easier to explain the clouds with the excessive velocity widths than those which have the same widths as ordinary galaxies. Unlike the cases of the Virgo clouds, which remain poorly explained as tidal debris, most of the Leo clouds are relatively close to major galaxies. This means they could indeed be unstable, transient features, as they would not have to survive for very long to reach their current separation from their parents. Cloud 6 is problematic in this regard  due to its isolation, but even here its line width is sufficiently low as to allow for a slow dispersal.

Applying heavy spectral smoothing to the data revealed that some of the clouds are more spatially extended than in the standard cube, but we did not find any evidence that they are connected to each other -- there does not appear to be any large-scale bridge connecting them, even for those clouds which are close in projection to each other. Although cloud 1 shows evidence of multiple kinematic components, spectral smoothing did not show this cloud had any significant extensions. While we cannot absolutely rule out that the clouds might be part of larger structure, in which case their measurements regarding the BTFR might have to be revised, we see no evidence for this in the data.

To summarise, there are a variety of possible explanations of the clouds, none of which can be definitively ruled out but none are entirely satisfying either :
\begin{itemize}
	\item All the clouds are tidal debris : some appear too isolated for this, and none of the nearest galaxies show the expected long \HI{} extensions.
	\item All the clouds are galaxian in nature though some are optically dark or very dim : there is no obvious reason why one cloud should be optically much brighter than the rest.
	\item Cloud 4, which alone has an optical counterpart, is a galaxy, while the rest are tidal debris : it seems an unlikely coincidence that cloud 4 should have such a qualitatively different nature to three other clouds in its vicinity with otherwise very similar quantitative \HI{} properties.
	\item If the clouds are tidal debris, some may have arisen in an interaction between M95 and M96 while some were produced by whatever process gave rise to the Leo Ring : the kinematics of the clouds do not match the velocity gradient of the Ring, and though M96 does show a warp in its \HI{} disc, neither it nor M95 show any larger-scale \HI{} emission predicted in numerical simulations of the formation of tidal debris.
\end{itemize}

In short, the clouds are consistent with both tidal and galaxian interpretations and neither scenario can be definitively ruled out. We are currently examining a variety of formation scenarios for the Leo Ring using numerical simulations, which may eventually shed light on these smaller but intriguing clouds.

\section*{Acknowledgments}
This work was supported by the Czech Ministry of Education, Youth and Sports from the large lnfrastructures for Research, Experimental Development and Innovations project LM 2015067, the Czech Science Foundation grant CSF 19-18647S, and the institutional project RVO 67985815.

RM acknowledges support of the NRAO. The National Radio Astronomy Observatory is a facility of the National Science
Foundation operated under cooperative agreement by Associated Universities, Inc.

This work is based on observations collected at Arecibo Observatory. The Arecibo Observatory is operated by SRI International under a cooperative agreement with the National Science Foundation (AST-1100968), and in alliance with Ana G. M\'{e}ndez-Universidad Metropolitana, and the Universities Space Research Association.

The SOFIA Science Center is operated by the Universities Space Research Association under NASA contract NNA17BF53C.

This research has made use of the NASA/IPAC Extragalactic Database (NED) which is operated by the Jet Propulsion Laboratory, California Institute of Technology, under contract with the National Aeronautics and Space Administration.

This work has made use of the SDSS. Funding for the SDSS and SDSS-II has been provided by the Alfred P. Sloan Foundation, the Participating Institutions, the National Science Foundation, the U.S. Department of Energy, the National Aeronautics and Space Administration, the Japanese Monbukagakusho, the Max Planck Society, and the Higher Education Funding Council for England. The SDSS Web Site is http://www.sdss.org/.

The SDSS is managed by the Astrophysical Research Consortium for the Participating Institutions. The Participating Institutions are the American Museum of Natural History, Astrophysical Institute Potsdam, University of Basel, University of Cambridge, Case Western Reserve University, University of Chicago, Drexel University, Fermilab, the Institute for Advanced Study, the Japan Participation Group, Johns Hopkins University, the Joint Institute for Nuclear Astrophysics, the Kavli Institute for Particle Astrophysics and Cosmology, the Korean Scientist Group, the Chinese Academy of Sciences (LAMOST), Los Alamos National Laboratory, the Max-Planck-Institute for Astronomy (MPIA), the Max-Planck-Institute for Astrophysics (MPA), New Mexico State University, Ohio State University, University of Pittsburgh, University of Portsmouth, Princeton University, the United States Naval Observatory, and the University of Washington.

This work has made use of the Digitized Sky Survey. The Digitized Sky Survey was produced at the Space Telescope Science Institute under U.S. Government grant NAG W-2166. The images of these surveys are based on photographic data obtained using the Oschin Schmidt Telescope on Palomar Mountain and the UK Schmidt Telescope. The plates were processed into the present compressed digital form with the permission of these institutions.

{}

\appendix
\section{Constructing the baryonic Tully Fisher relation}
In this paper we have shown three different version of the baryonic Tully Fisher relation. In figure \ref{fig:btfr} we use entirely our own data measurements, of both the AGES \HI{} data and the optical photometry from SDSS data, using either W50 or W20 to estimate rotation width (the two panels are otherwise identical). In figure \ref{fig:btfrmcg} we attempt to match our data to that described by \cite{btfr12}, which requires significant extra corrections. It is a legitimate question as to which best fit to the data is the \textit{best} best fit, as this can affect whether the optically dark clouds really do match the relation obtained from optically bright galaxies or not. We therefore describe the procedure for figure \ref{fig:btfrmcg} here in full.

\subsection{Distances}
In all previous AGES papers we generally estimate the distance by assuming galaxies are in pure Hubble flow with $H_{0}$\,=\,71\,\kms{}\,Mpc$^{-1}$. We use the measured heliocentric velocity to obtain distance via $d = v_{hel} / H_{0}$. However, \cite{aasdss}, who are concerned with accurate measurements for the BTFR, recommend using the estimation of \cite{aa100} : $d = v_{CMB} / H_{0}$ with $H_{0}$\,=\,70\,\kms{}\,Mpc$^{-1}$. 

We convert our measured $v_{hel}$ into $v_{CMB}$ using the NED velocity correction calculator. Converting RA and Dec to galactic longitude $l$ and latitude $b$ :
\[v_{CMB} = v_{hel} + v_{apex}(\,sin(b)sin(b_{apex}) + cos(b)cos(b_{apex})cos(l - l_{apex})\,) \]
Where, from \cite{fix}, $v_{apex} =\,$371\,\kms{}, $l_{apex} =\,$264.14 degrees, and $b_{apex} =\,$+48.26 degrees.

We use this distance estimation for all galaxies with $v_{hel}$\,$>$\,3,000\,\kms{}, i.e. beyond the Virgo cluster (\citealt{aa100} use this only for galaxies with $v_{hel}$\,$>$\,6,000\,\kms{}, with those at lower velocities having distance computed via a more complicated flow model). For galaxies below this redshift, our sample consists entirely of objects either in the Virgo cluster or Leo group, for which redshift-independent distance estimates are available. In Virgo, following \cite{me12}, galaxies are assigned to distances of either 17, 23 or 32\,Mpc depending on their spatial position, while in Leo, as described here, we assume a distance of 11.1\,Mpc for all objects.

\subsection{Inclination angles}
In \cite{me13} we used our own estimates of the inclination angle when correcting the line widths and we use these values here in figure \ref{fig:btfr}. However, to ensure consistency, for figure \ref{fig:btfrmcg} we use the values provided in the SDSS by the `expAB\_r' parameter as recommended by \citealt{aasdss}. We use this both for corrections to the velocity width and internal extinction, described below.

\subsection{Atomic mass}
The total mass of \HI{} is computed as in the main text via the equation :
\[M\HI{} = 2.36\times10^{5}\times d^{2}\times S_{\HI{}}\]
Where $S_{HI}$ is the total \HI{} flux and $d$ is the distance in Mpc as described above. For figure \ref{fig:btfrmcg}, correction for helium to get the mass of atomic gas is done by a simple multiplicative factor : M$_{gas} = 1.33 M_{\HI}$ following \cite{btfr12}, while for figure \ref{fig:btfr} we use a factor 1.36 for consistency with \cite{me13}.

\subsection{Stellar and baryonic mass}
Stellar masses are computed using our own measurements of $g$ and $i$ band optical photometry from the SDSS. We use the apparent magnitudes m$_{g,i}$ given in \cite{me12}. We correct these for galactic extinction to give \[m_{E\,g,i}\,=\,m_{g,i} - A_{g,i}\] where the attenuation A$_{g,i}$ is obtained from \cite{exschlaf}. Note that \cite{aasdss} favour the galactic extinction model of \citealt{exeshelg}, however towards Virgo the choice makes little difference as extinction here is anyway very low.

The extinction-corrected magnitudes are then converted to absolute magnitudes via the standard equation :
\[M_{E g,i} = -(\,5\,log(d) -5 - m_{E g,i}\,)\]
Where $d$ is the distance in parsecs. For figure \ref{fig:btfr}, following \cite{me13}, no additional correction for internal extinction is made. For figure \ref{fig:btfrmcg} we correct for internal extinction using the method described in \citealt{aasdss}. First, we generate a correction factor $\gamma$ which is zero if M$_{E} >$\,17.0, otherwise :
\[\gamma_{g} = -0.35M_{E g} - 5.95,\; \;  M_{E g} < -17.0 \]
\[\gamma_{i} = -0.15M_{E i} - 2.55,\; \;  M_{E i} < -17.0 \]
This is converted to the attenuation A :
\[A_{g,i} = \gamma_{g,i}log(b/a) \]
Where $b/a$ is the `expAB\_r' parameter obtained above. We then get our final values for the absolute magnitudes via :
\[M_{g,i} = M_{E g,i} + A_{g,i}\]
To convert to stellar mass we require the $i$ band luminosity in solar units, which is obtained from :
\[L_{i} = 10.0^{\frac{M_{i} - M_{i,\odot}}{-2.5}} \]
Where M$_{i,\odot}$ is the absolute magnitude of the Sun at $i$ band; following \citealt{aasdss} we use 4.58 whereas in our earlier papers (i.e. figure \ref{fig:btfr}) we have used 4.48. We can now convert to stellar mass, again following \citealt{aasdss} :
\[ M_{\ast} = 10.0^{ -0.68 + 0.70(g - i) + log(L_{i})}\]
While \citealt{aasdss} rely primarily on $g$ and $i$ band photometry as we do, they also give prescriptions to convert to stellar masses based on WISE IR photometry. They give two different methods to estimate the stellar mass in this way. We have found that their `McGaugh' formulation gives a somewhat better agreement with the BTFR of \cite{btfr12} and therefore we convert our stellar masses to this approximation. \citealt{aasdss} do not give a direct conversion between the stellar mass obtained above (the `Taylor' method in their nomenclature) and the `McGaugh' method, however they provide a conversion between the  `Taylor' stellar masses to the estimates of the GALEX-SDSS-WISE Legacy Catalog 2 :
\[ log(M_{\ast, GSWLC2}) = 1.052\,log(M_{\ast, Taylor}) - 0.369\]
Which in turn can be converted to the final stellar masses we use here :
\[ log(M_{\ast, McGaugh}) = (log(M_{\ast, GSWLC2}) + 0.975) / 1.084\]
Converting to linear units, the baryonic mass is then given simply by $M_{bary} = M_{gas} + M_{\ast, McGaugh}$.

\subsection{Line widths}
For figure \ref{fig:btfr} line widths are converted to circular rotation simply by taking half of the inclination-corrected value :
\[ V_{circ} = W20,50 / (2\,sin(i))\]
For figure \ref{fig:btfrmcg} we employ additional corrections. Based on \cite{btfr12}'s recommendation when using line width data, we only use the W20. We correct for finite spectral resolution following \cite{spring}. First, we calculate the S/N ratio as SNR = $(Flux_{peak} - rms) / rms$ (note the usual AGES peak S/N values are simply $(Flux_{peak} / rms$). We then compute a correction factor $\lambda$ based on SNR and the velocity resolution $v_{res}$ (10\,\kms{} for AGES) :
\[\lambda = 0.037\,v_{res} - 0.18,\; \; log(SNR) < 0.6  \]
\[\lambda =(0.0527\,v_{res} - 0.732) + (-0.027 v_{res} + 0.92)log(SNR),\; \; 0.6 \leq log(SNR) < 1.1  \]
\[\lambda = 0.023\,v_{res} + 0.28,\; \; log(SNR) \geq 1.1  \]
This then lets us calculate the velocity width correction factor :
\[\Delta S = 2 v_{res}\lambda  \]
Which we subtract linearly from the velocity width W50 or W20. We also correct for cosmological broadening, giving a final corrected velocity width of:
\[W_{corr} = (W20,50 - \Delta S) / (1 + z)  \]
This is then converted to a rotation velocity as above :
\[ V_{circ} = W_{corr} / (2\,sin(i))\]
\cite{btfr12} fit the BTFR using resolved rotation curves, but also provide an empirical correction factor to convert line width measurements to more accurate rotation velocities. This arises as line widths tend to represent the peak rotation velocity, which may be slightly higher than the flat part of the rotation curve. If M$_{gas} > $M$_{\ast}$, no additional correction is applied. If M$_{gas} < $M$_{\ast}$ :
\[ V_{circ,corr} = V_{circ} / 1.1\]
This is the final value for the velocity width used in figure \ref{fig:btfrmcg}. While \cite{spring} discuss possible additional corrections for turbulence, we have not applied this. Their favoured correction of linearly subtracting 6.5\,\kms{} is not appropriate for low line width objects as this can cause very large shifts and thus is strongly dependent on minor errors (discussed below). In addition, the clouds which form the focus of the present work are conceivably dominated by turbulent motions, so trying to correct for this would be a strange choice - to have a fair comparison, it is far simpler not to correct for turbulence in either the galaxies or the clouds.

\subsection{Sample selection and outliers}
For figure \ref{fig:btfr} we place only minimal restrictions on the sample : $i > 45$ degrees and \HI{}$_{def} < 0.6$. The inclination cut is to avoid large errors in the velocity width due to small errors in the inclination measurement. The deficiency cut is because, as we discuss in \cite{me13}, at high deficiencies we find galaxies with very low line widths even using the W20 line width. The reason for this is likely astrophysical, with the galaxies so stripped of gas that their remaining \HI{} no longer fully probes their rotation curve. As our interest here is in comparing the Leo clouds to the BTFR of normal galaxies, we exclude these unusual outliers from the plots.

For figure \ref{fig:btfrmcg} the emphasis is on accurately reproducing the BTFR of \cite{btfr12} and we impose an additional cut : peak $S/N > 10.0$. Equivalently this means the S/N level at the flux level of the W20 measurement is 2$\sigma$, which ensures a minimal inaccuracy due to noise. This cut is crucial. Without it, there is a very much larger scatter than in the figure shown here. That there is anyway a larger scatter in the line width data than \cite{btfr12}'s resolved rotation curve data is, as he discusses, unsurprising, and that most of the AGES galaxies lie within 2$\sigma$ scatter of this relation is probably about as good as we can expect. 

The peak 10$\sigma$ cut is somewhat arbitrary, but as a selection criteria it is broadly consistent with the Leo clouds we examine here. By gradually reducing this cut, we find that we do not see significantly more or stronger outliers until we reach 6$\sigma$. All of the Leo exceed 9$\sigma$. The Virgo clouds are a little fainter. The faintest is at 5.2$\sigma$, a level at which significantly more scatter is seen. However the next faintest is at 6.5$\sigma$, above the threshold where we see additional outliers, and the rest exceed 7$\sigma$. Therefore the choice of S/N cut for the plot does not greatly influence our comparison of the clouds with regard to the BTFR established for optically detected galaxies.

However, a \textit{few} more significant outliers remain apparent in our normal galaxy sample. The two galaxies with surprisingly low velocity widths (AGESVC1 251 and 252) may be astrophysically interesting, but they are both faint sources with low measured line widths and any errors in the inclination angle here may be significant. Two other galaxies (AGESVC1 230 and 288) have excessively high velocity widths. In both cases, examination of the spectra reveals peaks in the noise which just exceed the flux at 20\% of the peak value, skewing the W20 estimate. When we correct for this, both galaxies actually lie within the 2$\sigma$ scatter : if we apply the additional cleaning procedures (fitting a second-order polynomial to the whole data set) described in section \ref{sec:obspartone}, the noise peaks are pushed below the 20\% level and the measured width values are substantially reduced. This procedure was not used in the original published VC1 and VC2 data as it was not thought necessary; it was implemented later as a way of addressing much more egregious problems due to changes in the data reduction software. In a future AGES data release we may consider re-measuring the parameters using homogenous data processing but this is beyond the scope of the current work, and would be unfair to retroactively remove selected outliers based on their deviation from the BTFR when this is the very thing we are trying to examine. Note that the optically dark clouds in Virgo also all have follow-up observations.

\subsection{Comparison of the different BTFRs}
The different procedures adopted for the BTFR plots here cause substantial changes in slopes; exactly as discussed in \cite{manybtfr, btfr12} this changes the slope (the logarithm of the exponent) from 3 to 4. At $V_{circ} = 100$\,\kms{}, given that this also changes the normalisation, this is equivalent to a baryonic mass change of about 0.5 dex. \cite{manybtfr, btfr12} choose the fit which gives a slope of 4, arguing that since this also gives the lowest scatter this is more likely to be the correct value. Given the many different choices available when correcting the various aspects of the plot, we refrain from commenting on which is preferable; our intention here is to compare the clouds with established BTFRs rather than advocate for any particular methodology. In any case, the basic result that some clouds clearly deviate from the BTFR while some are within the general scatter appears to be robust.

The systematic effects in fitting the BTFR are the dominant source of error. For this reason we avoid showing error bars in the plots as the formal measurement errors are misleadingly small. In the case of the Leo clouds the W20 errors are 5-15\,\kms{}, allowing for a maximum change in width of 0.1 dex (but note we account for spectral smoothing above and see the discussion below). The error in the measured \HI{} flux is typically around 20\%, allowing for an shift in mass of about 0.07 dex. Such shifts could in principle combine to move the clouds further from the \cite{btfr12} relation, but it seems an unlikely coincidence that five features have been found which all lie within the observed scatter. Still, the only way to be certain would be deep interferometric follow-up.

Finally, another concern is whether the comparison of line widths from demonstrably rotating galaxies and the motions in clouds of unknown nature is physically meaningful. For the sake of homogeneity we have applied the same correction for rotation (i.e. halving the line width) for the clouds as for the galaxies, though we cannot correct their measurements for inclination. Therefore if they are rotating systems their velocities plotted here are lower limits, meaning they might actually be outliers from the BTFR after all. However, any rotating system which deviates from the BTFR in this way would seem to be interesting almost by definition. Conversely, there is no particular reason to expect turbulence-dominated clouds to follow the usual BTFR. On this basis, we argue that the comparison is an interesting one regardless of the nature of the kinematics of the different objects. 


\begin{thebibliography}{}

\bibitem[\protect\citeauthoryear{Adams et al.}{2015}]{adamsclouds} 
Adams E. K. et al., 2015, A\&A, 580, 134 
\bibitem[\protect\citeauthoryear{Auld et al.}{2006}]{auld}
Auld R., Minchin R. F., Davies J. I., et al. 2006, MNRAS, 371, 1617
\bibitem[\protect\citeauthoryear{Bekki et al.}{2005}]{bekki}
Bekki K., Koribalski B. S., Kilborn V. A., 2005, MNRAS, 363, 21
\bibitem[\protect\citeauthoryear{Bellazzini et al.}{2018}]{bellazz}
Bellazzini M.,  Armillotta L., Perina S., Magrini L., Cresci G., Beccari G., Battaglia G., Fraternali F., de Zeeuw P. T., Martin N. F., Calura F., Ibata R., Coccato L., Testa V., Correnti M., 2018, MNRAS, 476, 4565
\bibitem[\protect\citeauthoryear{Calura, Bellazzini \& D'Ercole}{2020}]{secco}
Calura F., Bellazzini M., D'Ercole A.,2020, MNRAS, 499, 5873
\bibitem[\protect\citeauthoryear{Corbelli et al.}{2021}]{leomet}
Corbelli E., Cresci G., Mannucci F.; Thilker D., Venturi G., 2021, 908, 39
\bibitem[\protect\citeauthoryear{Davies et al.}{2004}]{d04}
Davies J., Minchin R., Sabatini S., van Driel W., Baes M., Boyce P., de Blok W. J. G., Disney M., 2004, MNRAS, 349, 922
\bibitem[\protect\citeauthoryear{Davies et al.}{2011}]{d11}
Davies J. I., Auld R., Burns L., Minchin R., Momjian E., Schneider S., Smith M., Taylor R., van Driel W., 2011, MNRAS, 415, 1883
\bibitem[\protect\citeauthoryear{Durbala et al.}{2020}]{aasdss}
Durbala A., Finn R.A., Crone Odekon M., Haynes M.P., Koopmann R.A., O'Donohue A.A., 2020, Astron. J. 160, 721
\bibitem[\protect\citeauthoryear{Duc \& Bournaud}{2008}]{duc}
Duc P. A., Bournaud F., 2008, ApJ, 673, 787
\bibitem[\protect\citeauthoryear{Ferguson \& Sandage}{1990}]{fergie}
Ferguson H. C., Sandage A., 1990, AJ, 100, 1
\bibitem[\protect\citeauthoryear{Fixsen et al.}{1996}]{fix}
Fixsen D. J., Cheng E. S., Gales J. M., Mather J. C., Shafer R. A., Wright, E. L., 1996, ApJ, 473, 576
\bibitem[Giovanelli et al. (2005)]{alfalfa} Giovanelli R., Haynes M. P., Kent B. R., Perillat P., Saintonge A., Brosch N., Catinella B., Hoffman G. L., et al., 2005, AJ, 130, 2598
\bibitem[\protect\citeauthoryear{Guo et al.}{2020}]{guo}
Guo Q., Hu H., Zheng Z., et al., 2020, NatAs, 4, 246
\bibitem[\protect\citeauthoryear{Haynes, Giovanelli \& Kent}{2007}]{aavhi21}
Haynes M. P., Giovanelli R., Kent B., 2007, ApJ, 665, 19
\bibitem[\protect\citeauthoryear{Haynes et al.}{2018}]{aa100}
Haynes M. P., Giovanelli R., Kent, B. R., et al., 2018, ApJ, 861, 49
\bibitem[\protect\citeauthoryear{J\'{o}zsa et al.}{2022}]{jozsa}
J\'{o}zsa G. I. G., Jarrett T. H., Cluver M. E., et al., 2022, ApJ, 926, 167
\bibitem[\protect\citeauthoryear{Karachentsev \& Karachentseva}{2004}]{kk}
Karachentsev I. D., Karachentseva V. E., 2004, ARep, 48, 267
\bibitem[\protect\citeauthoryear{Keenan et al.}{2016}]{olivia}
Keenan O. C., Davies J. I., Taylor R., Minchin R. F., 2016, MNRAS, 456, 951
\bibitem[\protect\citeauthoryear{Kilborn et al.}{2006}]{kilborn}
Kilborn V. A., Forbes D. A., Koribalski B. S., Brough S., Kern K., 2006, MNRAS, 371, 739
\bibitem[\protect\citeauthoryear{Leisman et al.}{2016}]{aaleoleis}
Leisman L., Haynes M.P., Giovanelli R., Józsa G., Adams E.A.K., Hess K.M. 2016, MNRAS 463, 1692
\bibitem[\protect\citeauthoryear{Lelli, McGaugh \& Schombert}{2016}]{lelli}
Lelli F., McGaugh S. S., Schombert J. M., 2016, ApJ, 816, 14
\bibitem[\protect\citeauthoryear{Leroy et al.}{2008}]{leroy}
Leroy A. K., Walter F., Brinks E., Bigiel F., de Blok W. J. G., Madore B., Thornley M. D., 2008, AJ, 136, 2782
\bibitem[\protect\citeauthoryear{McGaugh et al.}{2000}]{btfr}
McGaugh S. S., Schombert J. M., Bothun G. D., de Blok W. J. G., 2000, ApJ, 533, 99
\bibitem[\protect\citeauthoryear{McGaugh}{2005}]{manybtfr}
McGaugh S. S., 2005, ApJ, 632, 859
\bibitem[\protect\citeauthoryear{McGaugh}{2012}]{btfr12}
McGaugh S. S., AJ, 143, 40
\bibitem[\protect\citeauthoryear{Michel-Dansac et al.}{2010}]{leocollide}
Michel-Dansac L., Duc P-A., Bournaud F., Cuillandre J-C., Emsellem E., Oosterloo T., Morganti R., Serra P., Ibata R., 2010, ApJ, 717, 143
\bibitem[\protect\citeauthoryear{Mihos et al.}{2012}]{mihosm101}
Mihos J. C., Keating L. M., Holley-Bockelmann K., Pisano D. J., Kassim N. E., 2012, ApJ, 761, 186
\bibitem[\protect\citeauthoryear{Minchin et al.}{2007}]{m07}
Minchin R., Davies J. I., Disney M., Grossi M., Sabatini S., Boyce P., Garcia D., Impey C., et al., 2007, ApJ, 670, 1056
\bibitem[\protect\citeauthoryear{Minchin et al.}{2010}]{m10}
Minchin R. F., Momjian E., Auld R., Davies J. I., Valls-Gabaud D., Karachentsev I. D., Henning P. A., O'Neil K. L., et al., 2010, MNRAS, 140, 1093
\bibitem[\protect\citeauthoryear{M{\"u}ller, Jerjen, \& Binggeli}{2018}]{mull}
M{\"u}ller O., Jerjen H., Binggeli B., 2018, A\&A, 615, 105
\bibitem[\protect\citeauthoryear{Oosterloo et al.}{2005}]{oo05}
Oosterloo T., van Gorkom J., 2005, A\&A, 437, 19
\bibitem[\protect\citeauthoryear{Oosterloo, Heald \& de Blok}{2013}]{oo13}
Oosterloo T. A., Heald G. H., de Blok W. J. G, 2013, A\&A, 555, 7
\bibitem[\protect\citeauthoryear{Mancera Piña et al.}{2020}]{pina}
Mancera Piña P. E., Fraternali F., Oman K. A., et al., 2020, MNRAS, 495, 3636
\bibitem[\protect\citeauthoryear{Mancera Piña et al.}{2021}]{pina22}
Mancera Piña P. E., Fraternali F., Oosterloo T., Adams E. A. K., Oman K. A., Leisman L, 2022, MNRAS, 512, 3230
\bibitem[\protect\citeauthoryear{Putman et al.}{2002}]{put}
Putman M. E., de Heij V., Staveley-Smith L., Braun R., Freeman K. C., Gibson B. K., Burton W. B., Barnes D. G., et al., 2002, AJ, 123, 873
\bibitem[\protect\citeauthoryear{Román et al.}{2021}]{fadug}
Román J., Jones M. G., Montes M., Verdes-Montenegro L, Garrido J., Sánchez S., 2021, A\&A, 649, 14
\bibitem[\protect\citeauthoryear{Rosenberg et al.}{2014}]{rose}
Rosenberg J. L., Haislmaier K., Giroux M. L., Keeney B. A., Schneider, S. E., 2014, ApJ, 740, 64
\bibitem[\protect\citeauthoryear{Saintonge}{2007}]{aasn}
Saintonge A., 2007, AJ, 133, 2087
\bibitem[\protect\citeauthoryear{Schneider}{1989}]{schnleo}
Schneider, S. E. 1989, ApJ, 343, 94
\bibitem[\protect\citeauthoryear{Sil'chenko et al.}{2003}]{leoprim}
Sil'chenko O. K.,  Moiseev A. V., Afanasiev V. L., Chavushyan V. H., Valdes J. R., 2003, ApJ, 591, 185
\bibitem[\protect\citeauthoryear{Stierwalt et al.}{2009}]{aaleo}
Stierwalt S., Haynes M. P., Giovanelli R., Kent B. R., Martin A. M., Saintonge A., Karachentsev I. D., Karachentseva V. E., 2009, AJ, 138, 338
\bibitem[\protect\citeauthoryear{Sault, Teuben, \& Wright}{2005}]{miriad}
Sault R.J., Teuben P.J., Wright M.C.H., 1995, Astronomical Data Analysis Software and Systems IV, ed. R. Shaw, H.E. Payne, J.J.E. Hayes, ASP Conference Series, 77, 433-436.
\bibitem[\protect\citeauthoryear{Schlafly \& Finkbeiner}{2011}]{exschlaf}
Schlafly E. F., Finkbeiner D. P., ApJ, 737, 103
\bibitem[\protect\citeauthoryear{Schlegel, Finkbeiner \& Davis}{1998}]{exeshelg}
Schlegel D. J., Finkbeiner D. P., Davis M., 1998, ApJ, 500, 525
\bibitem[\protect\citeauthoryear{Springob et al.}{2005}]{spring}
Springob C. M., Haynes M. P., Giovanelli R., Kent B. R., 2005, ApJ, 160, 149
\bibitem[\protect\citeauthoryear{Taylor et al.}{2012}]{me12}
Taylor R., Auld R., Davies J. I., Minchin R. F., 2012, MNRAS, 423, 787
\bibitem[\protect\citeauthoryear{Taylor et al.}{2013}]{me13}
Taylor R., Minchin R. F., Herbst H., Davies J. I., Rodriguez R., Vasquez C., 2013, MNRAS, 428, 459
\bibitem[\protect\citeauthoryear{Taylor et al.}{2014}]{me14}
Taylor R., Minchin, R. F., Herbst H., Smith R., 2014, MNRAS, 442, 46
\bibitem[\protect\citeauthoryear{Taylor}{2015}]{me15}
Taylor R., 2015, A\&C, 13, 67
\bibitem[\protect\citeauthoryear{Taylor et al.}{2016}]{me16}
Taylor R., Davies J. I., J\'{a}chym P., Keenan O., Minchin R. F., Palou\v{s} J., Smith R., W\"{u}nsch R., 2016, MNRAS, 461, 300
\bibitem[\protect\citeauthoryear{Taylor et al.}{2017}]{me17}
Taylor R., Davies J. I., J\'{a}chym P., Keenan O., Minchin R. F., Palou\v{s} J., Smith R., W\"{u}nsch R., 2017, MNRAS, 467, 3648
\bibitem[\protect\citeauthoryear{Taylor et al.}{2020}]{me20}
Taylor R., Köppen J, Jáchym P.,  Minchin R., Palouš J., Wünsch R., 2020, AJ, 159, 218
\bibitem[\protect\citeauthoryear{Thilker et al.}{2009}]{thileo}
Thilker D. A., Donovan J., Schiminovich D., Bianchi L., Boissier S., Gil de Paz A., Madore B. F., Martin D. C., Seibert M., 2009, Nature, 457, 990
\bibitem[\protect\citeauthoryear{Tonnesen \& Bryan}{2010}]{ton10}
Tonnesen S., Bryan G. L., 2010, ApJ, 709, 1203
\bibitem[\protect\citeauthoryear{Toomre \& Toomre}{1972}]{tooms}
Toomre A., Toomre J., 1972, ApJ, 178,623
\bibitem[\protect\citeauthoryear{van Dokkum et al.}{2018}]{vd}
van Dokkum P., Danieli S., Cohen Y., Merritt A., Romanowsky A. J.,
Abraham R., Brodie J., Conroy C., et al. 2018, Nature, 555, 629
\bibitem[\protect\citeauthoryear{Watkins et al.}{2014}]{watkins}
Watkins A. E., Mihos J. C., Harding P., Feldmeier J. J., 2014, ApJ, 791, 38
\bibitem[\protect\citeauthoryear{Wolfe et al.}{2013}]{wolfe}
Wolfe S. A., Pisano D. J., Lockman F. J., McGaugh S. S, Shaya E. J., 2013, Nature, 497, 224
\bibitem[\protect\citeauthoryear{Wong et al.}{2021}]{wong}
Wong O. I., Stevens A. R. H., For B-Q., 2021, MNRAS, 507, 2905
\bibitem[\protect\citeauthoryear{Xi et al.}{2021}]{auds}
Xi H., Staveley-Smith L., For B-Q., Freudling W., Zwaan M., Hoppmann L., Liang F-U., Peng B,, 2021, MNRAS, 501, 4550
\bibitem[\protect\citeauthoryear{Yu, Ho, \& Wang.}{2020}]{yu20}
Yu N., Ho L. C., Wang J., 2020, ApJ, 898, 102
\bibitem[\protect\citeauthoryear{Yu et al.}{2022}]{yu22}
Yu N., Ho L. C., Wang J., Li H., ApJ, in press
\end{thebibliography}
\end{document}